\documentclass[sigconf]{acmart}
\usepackage{multirow}

\AtBeginDocument{%
  }

\copyrightyear{2026}
\acmYear{2026}
\setcopyright{cc}
\setcctype{by-nc-nd}
\acmConference[CHI EA '26]{Extended Abstracts of the 2026 CHI Conference on Human Factors in Computing Systems}{April 13--17, 2026}{Barcelona, Spain}
\acmBooktitle{Extended Abstracts of the 2026 CHI Conference on Human Factors in Computing Systems (CHI EA '26), April 13--17, 2026, Barcelona, Spain}
\acmDOI{10.1145/3772363.3798444}
\acmISBN{979-8-4007-2281-3/2026/04}

\begin{document}

\title{ORCA: ORchestrating Causal Agent}

\author{Joanie Hayoun Chung}
\affiliation{%
  \institution{Department of Statistics,}
  \city{}
  \country{}}
\affiliation{%
  \institution{Korea University}
  \city{Seoul}
  \country{South Korea}}
\email{jchung02@korea.ac.kr}

\author{Sumin Lee}
\affiliation{%
  \institution{Department of Statistics,}
  \city{}
  \country{}}
\affiliation{%
  \institution{Korea University}
  \city{Seoul}
  \country{South Korea}}
\email{leesumin0924@korea.ac.kr}

\author{Sungbin Lim}
\thanks{Corresponding author. E-mail: \texttt{sungbin@korea.ac.kr}.}
\affiliation{%
  \institution{Department of Statistics}
  \city{}
  \country{}}
\affiliation{
  \institution{Korea University}
  \city{}
  \country{}
}
\affiliation{%
  \institution{LG AI Research}
  \city{Seoul}
  \country{South Korea}}
\email{sungbin@korea.ac.kr}

\renewcommand{\shortauthors}{Chung et al.}

\begin{abstract}
Causal analysis on relational databases is challenging, as analysis datasets must be repeatedly queried from complex schemas.
Recent LLM systems can automate individual steps, but they hardly manage dependencies across analysis stages, making it difficult to preserve consistency between causal hypothesis.
We propose ORCA (ORchestrating Causal Agent), an interactive multi-agent framework to enable coherent causal analysis on relational databases by maintaining shared state and introducing human checkpoints.
In a controlled user study, participants using ORCA successfully completed end-to-end analysis more often than with a baseline LLM (GPT-4o-mini) assistant by 42 percentage points, achieved substantially lower ATE error, and reduced time spent on repetitive data exploration and query refinement by 76\% on average.
These results show that ORCA improves both how users interact with the causal analysis pipeline and the reliability of the resulting causal conclusions.
\end{abstract}




\maketitle

\section{Introduction}
Understanding causality is essential in high-stakes domains such as medicine, public policy, and business strategy, where analytical errors can directly affect safety, equity, and resource allocation~\cite{kuhne2022causal, heckman2022econometric, bojinov2020importance}.
Reliable causal analysis requires both domain knowledge to articulate a plausible data-generating process and statistical expertise to select valid identification and estimation strategies. 
These requirements are intensified by large, relationally complex databases, where variables are distributed across multiple tables~\cite{dong2022high, berkessa2024review, mondal2025fast}.
Analysts must obtain datasets through repeated joins, aggregations, and variable definitions.
This iterative process is often time-consuming, and because these decisions determine the inputs to causal models, even small ambiguities in early dataset exploration can propagate across stages and distort downstream causal conclusions.
Recent LLM systems can automate parts of this workflow~\cite{maojun2025lambda, testini2025measuring, kulkarni2025agent}, particularly SQL generation and code execution~\cite{wu2024autogen}.
Yet they often fail to manage dependencies between data-related and modeling decisions across stages, leaving analysts without support for preserving consistency across the workflow.

Under these conditions, we propose ORCA (ORchestrating Causal Agent), an interactive multi-agent framework for reliable causal analysis on relational databases.
ORCA decomposes the workflow into specialized agents for data exploration, causal discovery, and effect estimation, coordinated by an Orchestrator that maintains cross-stage consistency and mediates user feedback.
Rather than fully automating decisions, ORCA introduces checkpoints that allow analysts to approve or revise artifacts. 
The main contributions of this paper are as follows: 
(1) We introduce ORCA, an interactive multi-agent system that supports the workflow of causal analysis on relational databases by guiding user interaction through explicit checkpoints.
(2) We demonstrate that ORCA improves causal outcome quality by enforcing cross-stage consistency, yielding more accurate graphs and ATE estimates.
(3) We present REEF, a semi-synthetic relational e-commerce benchmark with an explicit data-generating process, and evaluate ORCA via controlled experiments and a user study showing higher completion rates, lower ATE error, and reduced exploration time.

\section{ORCA: Orchestrating Causal Agent}
\label{sec:method}
\begin{figure*}[t]
    \centering
    \includegraphics[width=0.9\textwidth]{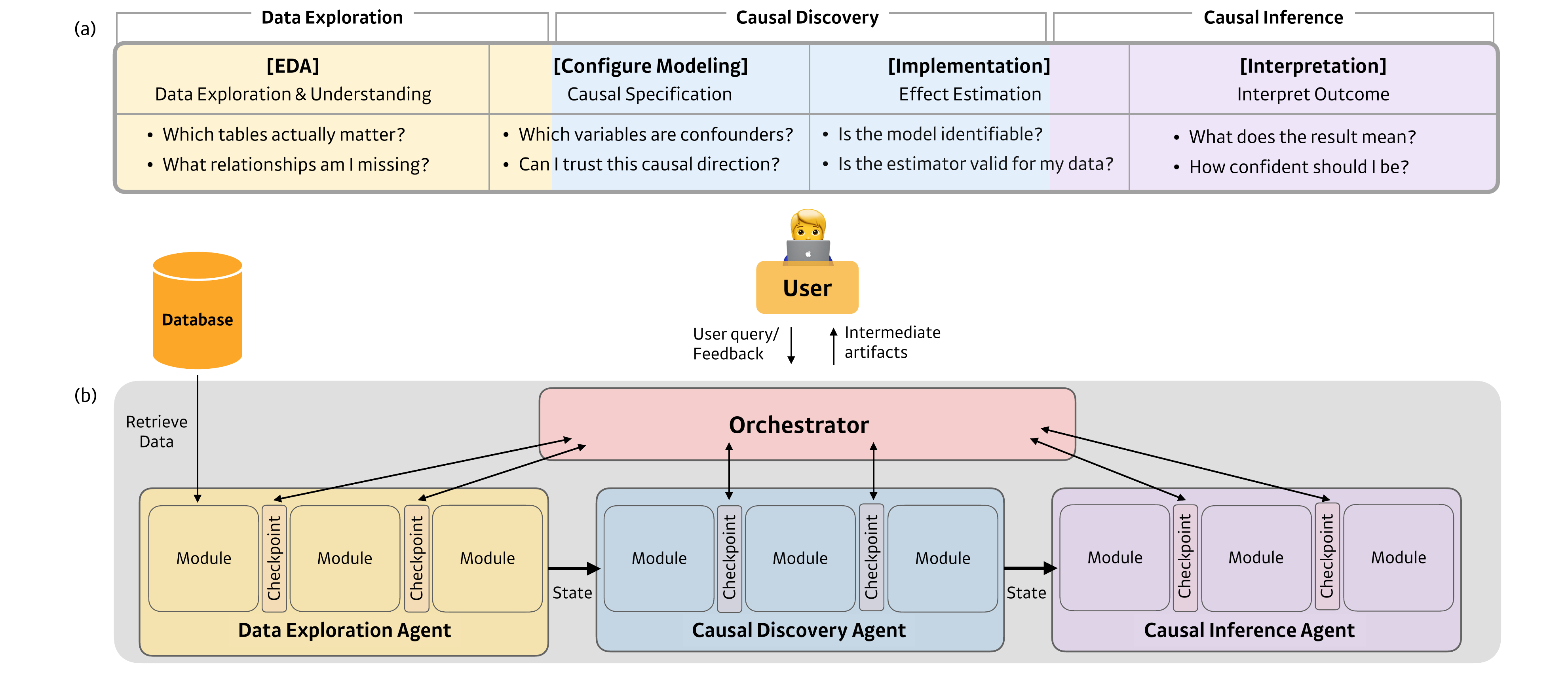}
  \caption{
  (a) The step-wise causal analysis procedure, highlighting the analytical questions that arise at each stage.
  (b) ORCA is aligned with this procedure by an orchestrated multi-agent framework.
  Three specialized agents share state of resulting artifacts from each module, while a central Orchestrator operates these agents sequentially during the analysis.
  User interaction is mediated by the Orchestrator through checkpoints where intermediate artifacts are exposed for inspection and feedback.}
  \Description{This figure presents an overview of the ORCA framework and its alignment with a step-wise causal analysis pipeline.
The top row illustrates four sequential stages of causal analysis: data exploration and understanding, causal specification, effect estimation, and interpretation. Each stage highlights representative analytical questions, such as identifying relevant variables, specifying confounders, selecting estimators, and interpreting uncertainty.
The bottom section depicts the ORCA system architecture. A central orchestrator coordinates three specialized agents: the Data Exploration Agent, the Causal Discovery Agent, and the Causal Inference Agent. These agents operate sequentially while maintaining shared state across stages. A user interacts with the orchestrator by providing queries and feedback at explicit checkpoints. Solid arrows indicate user interaction points where intermediate artifacts are exposed for inspection and revision before proceeding.}
  \label{fig:orca_overview_figure}
\end{figure*}

\subsection{Research Motivation and Framework Overview}
To investigate practical challenges in LLM-assisted causal analysis, we conducted a user study with twelve participants performing end-to-end causal tasks on a relational database using a GPT-4o-mini assistant with code execution tools.
Even with automated assistance, participants spent 77\% of their total analysis time on data exploration alone (see Figure~\ref{fig:user_study_result}(a)).
Repeated exploration made it difficult to track whether earlier intermediate artifacts remained valid, leading many participants to proceed without explicitly verifying these outputs.
This is critical in causal discovery and inference, since these methods are only applicable when the assumptions they rely on hold.
As a result, causal estimation was frequently applied to datasets that no longer satisfied the assumptions required for valid inference, yielding poor estimation outcomes.
Only 8.3\% of final ATE estimates fell within the ground-truth confidence interval (see Table~\ref{tab:user_study_subfigure}).
See Section~\ref{sec:user_study_results} for detailed quantitative and qualitative results.
These observations suggest that the core challenge lies not in automating individual steps, but in maintaining coherence across interdependent stages of causal analysis.
We formalize this as cross-stage consistency: the set of constraints requiring that (i) the relational join and aggregation decisions made during data exploration are reflected in the variable definitions passed to causal discovery, (ii) the causal graph produced by discovery encodes only assumptions that hold over the constructed dataset, and (iii) the identification strategy selected for effect estimation remains valid under those same assumptions.
Violations at any stage, such as changing a variable's aggregation level after a causal graph has been constructed can silently invalidate downstream estimates.

ORCA is designed to address this problem by structuring causal analysis as a sequence of stages with explicit artifacts and checkpoints.
It automates routine operations while preserving shared state and enabling users to inspect and revise intermediate results before errors propagate.
Figure~\ref{fig:orca_overview_figure} illustrates this stage-wise view and shows how ORCA realizes it through an orchestrated agent framework. 

The following subsections describe each specialist agent in terms of its responsibilities and checkpoints within the end-to-end causal workflow.
Implementation details are provided in Appendix~\ref{appendix:framework_detail} for reproducibility.

\subsection{Data Exploration Agent}
\begin{figure*}[t]
\centering
\includegraphics[width=0.95\linewidth]{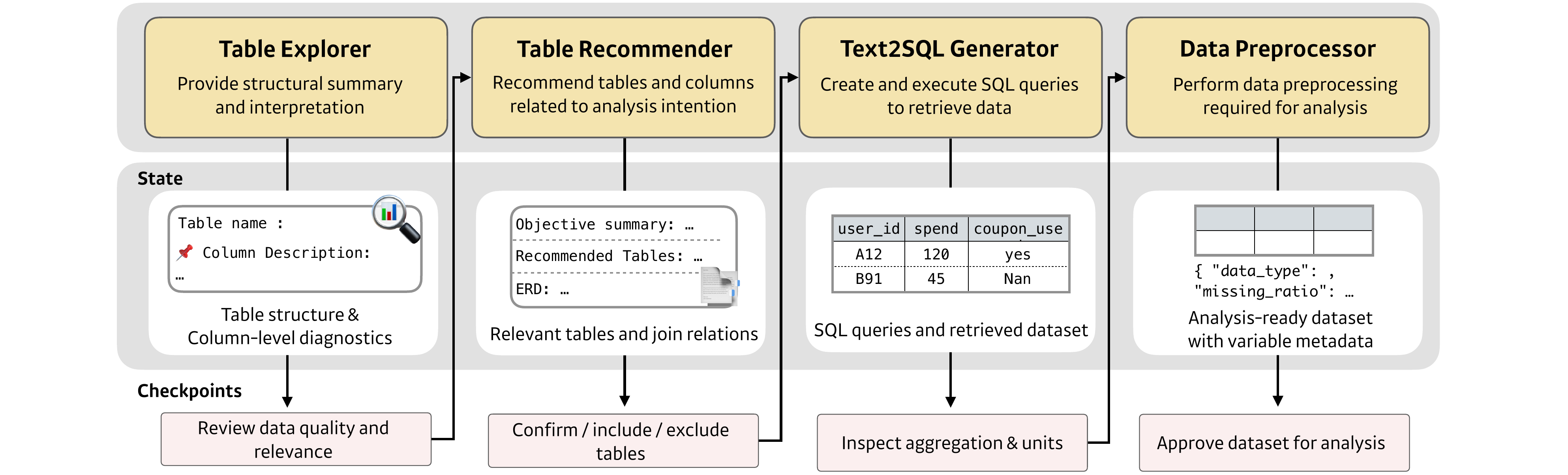}
\caption{ORCA’s Data Exploration Agent and its interaction structure. 
The agent prepares an analysis-ready dataset through four modules.
Each module produces an intermediate artifact saved in state (middle row) exposing table semantics, relational scope, query structure, and outcomes.
At each checkpoint (bottom row), users review and validate these outputs before proceeding.}
\Description{This figure illustrates the internal structure of ORCA’s Data Exploration Agent and its interaction with the user.
The pipeline consists of four modules arranged from left to right: Table Explorer, Table Recommender, Text2SQL Generator, and Data Preprocessor. Each module produces an intermediate artifact that captures progressively refined information, including table summaries, recommended tables and relationships, executed SQL queries, and an analysis-ready dataset with metadata.
Below the modules, explicit checkpoints are shown where the user reviews data quality, confirms or excludes tables, inspects aggregation levels and units, and approves the final dataset. The Orchestrator mediates transitions between modules and checkpoints, ensuring that user-validated artifacts are carried forward consistently.}
\label{fig:detail_exploration_figure}
\end{figure*}

The Data Exploration Agent supports preparation of analysis-ready datasets.
In relational databases, data exploration is a major bottleneck, involving repeated table lookup, interpretation of unfamiliar columns, and data quality assessment.
The agent decomposes this process into modules that leverage precomputed metadata and column-level statistics to guide exploration, producing artifacts that are validated at checkpoints and stored as shared state for consistency.

As illustrated in Figure~\ref{fig:detail_exploration_figure}, the Data Exploration Agent consists of four modules that can be used independently or sequentially.
(1) The \textit{Table Explorer} supports early-stage sense-making by summarizing table semantics, data quality, and relational context.
By integrating column-level statistics with schema metadata, it helps users assess whether a table is suitable for defining treatment or outcome variables.
(2) The \textit{Table Recommender} determines the relational scope of the analysis by identifying relevant tables and valid join paths aligned with the user’s analytical intent.
(3) The \textit{Text2SQL Generator} translates natural language requests into SQL queries and iteratively validates them for both syntactic and semantic alignment with the intended unit of analysis.
(4) Finally, the \textit{Data Preprocessor} applies lightweight transformations and annotates variable types and cardinalities, producing an analysis-ready dataset.

The effectiveness of these modules is evaluated on the REEF and BIRD benchmarks, demonstrating higher-quality table descriptions, more accurate table retrieval, and more reliable Text2SQL execution; detailed evaluation metrics and results are provided in Appendix~\ref{appendix:exploration-eval}.
Implementation-level details of the Data Exploration Agent, including module-specific prompts, inputs and outputs, and execution logic, are provided in Appendix~\ref{appendix:dataexploration_detail}.

\subsection{Causal Discovery Agent}
\begin{figure*}[t]
\centering
\includegraphics[width=0.95\linewidth]{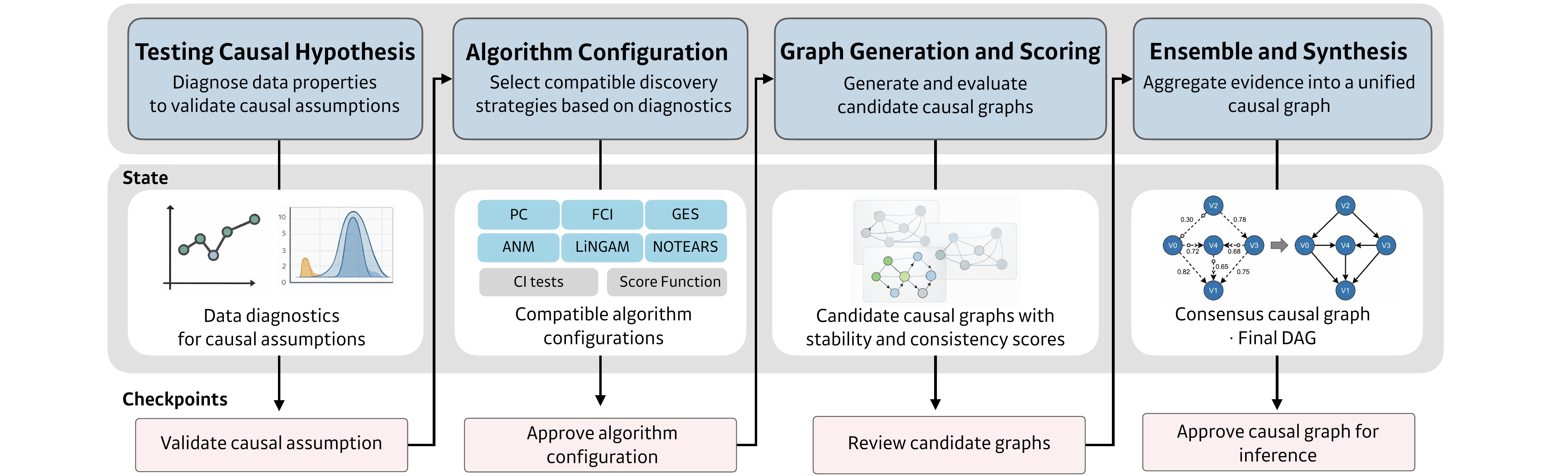}
\caption{ORCA’s diagnostics-driven Causal Discovery Agent with user checkpoints.
Starting from a preprocessed dataset, the agent evaluates causal assumptions, configures compatible discovery strategies, and generates candidate causal graphs.
Intermediate artifacts in state (middle row) expose diagnostics, configurations, and candidate graphs.
Checkpoints (bottom row) prompt users to validate assumptions, approve configurations, and review graph structures before inference.}
\Description{This figure depicts the diagnostics-driven workflow of ORCA’s Causal Discovery Agent.
Starting from a preprocessed dataset, the agent proceeds through four stages: testing causal hypotheses, algorithm configuration, graph generation and scoring, and ensemble synthesis. Each stage produces intermediate artifacts, including diagnostic plots for causal assumptions, compatible algorithm configurations, candidate causal graphs with stability scores, and a final consensus causal graph.
User checkpoints appear beneath each stage, allowing validation of assumptions, approval of algorithm choices, review of candidate graphs, and confirmation of the final causal graph before it is passed to causal inference. The Orchestrator manages state transitions and ensures that validated decisions are consistently applied.}
\label{fig:detail_dicovery_figure}
\end{figure*}

The Causal Discovery Agent aims to construct a reliable causal graph.
Causal discovery is inherently ill-posed~\cite{peters2017elements}, as multiple graph structures can be consistent with the same observational distribution, and no single modeling assumption can be fixed a priori~\cite{shimizu2006linear, zhang2012identifiability}.
Recent LLM-based approaches assist discovery by proposing structural hypotheses or incorporating human-provided cues~\cite{khatibi2024alcm}, but often rely on implicit assumptions not validated against data, leading to fragile or misleading graphs~\cite{jin2023can}. 
To address this, ORCA adopts a diagnostics-driven approach that evaluates assumption-relevant data properties and guides users to interpret and inspect the decisions and artifacts.

As illustrated in Figure~\ref{fig:detail_dicovery_figure}, the Causal Discovery Agent consists of four sequential modules with shared state.
(1) The \textit{Testing Causal Hypothesis} module evaluates whether key assumptions such as linearity and noise properties are supported by the data, constraining the applicable algorithm classes.
(2) The \textit{Algorithm Configuration} module maps these diagnostics to compatible algorithm families and requests user approval.
(3) The \textit{Graph Generation and Scoring} module generates candidate causal graphs using the configured algorithms and compares them using stability and consistency criteria, enabling user inspection of alternative structures.
(4) Finally, the \textit{Ensemble and Synthesis} module aggregates evidence across candidates to output both an uncertainty-preserving graph for inspection and a fully directed acyclic graph for inference.
Users review the synthesized graph before used for causal estimation.

Across synthetic settings, the agent yields more stable causal graphs, consistently reducing structural error (SHD) in scale-free settings (See Appendix~\ref{appendix:discovery-eval} for full experimental results).
Appendix~\ref{appendix:causaldiscovery_detail} specifies the procedures, diagnostic tests, thresholds, scoring definitions, and synthesis algorithms used.

\subsection{Causal Inference Agent}
\begin{figure*}[t]
\centering
\includegraphics[width=0.95\linewidth]{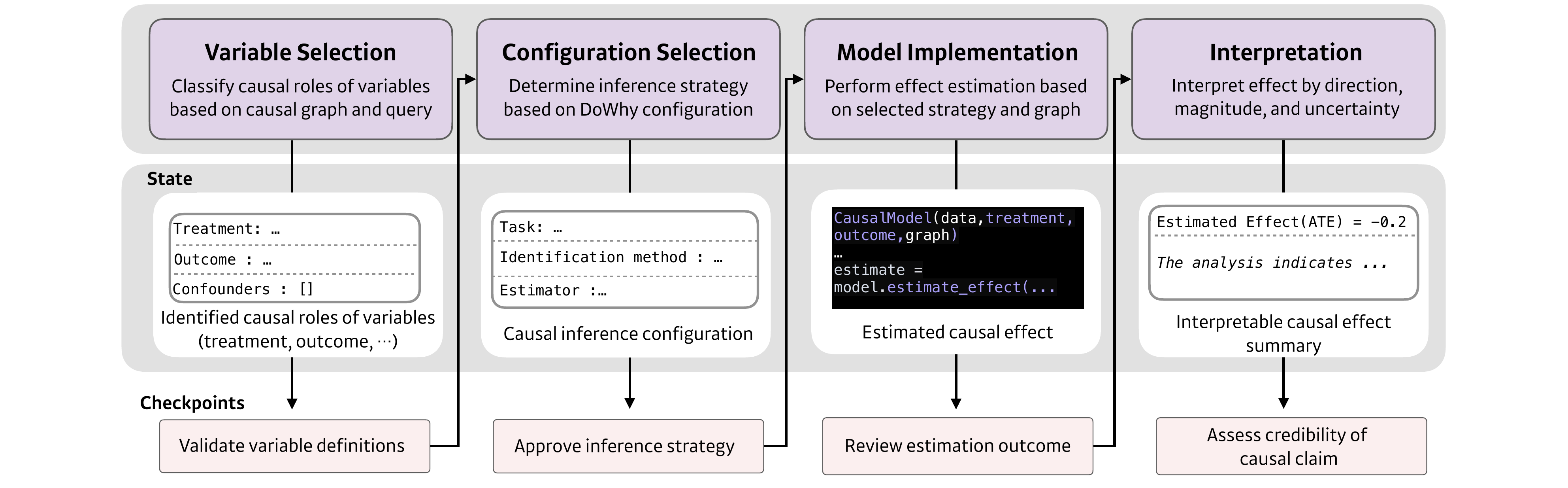}
\caption{ORCA’s Causal Inference Agent.
Given a causal graph and an analysis-ready dataset, the agent selects inference strategy, executes effect estimation, and interprets results.
Intermediate artifacts in state (middle row) expose variable roles, inference strategy, and estimation outputs and at each checkpoint (bottom row), users validate decisions and assess the results.}
\Description{This figure shows the workflow of ORCA’s Causal Inference Agent, which estimates and interprets causal effects given a causal graph and an analysis-ready dataset.
The pipeline consists of four stages: variable selection, configuration selection, model implementation, and interpretation. Intermediate artifacts include identified causal roles of variables (treatment, outcome, confounders), selected inference strategies and estimators, estimated causal effects, and an interpretable summary describing effect direction, magnitude, and uncertainty.
At each stage, user checkpoints allow validation of variable definitions, approval of inference strategies, review of estimation results, and assessment of the credibility of the causal claim. The Orchestrator ensures that only validated artifacts are propagated across stages.}
\label{fig:detail_inference_figure}
\end{figure*}

The Causal Inference Agent supports reliable effect estimation through step-wise inference with checkpoints and dependency-aware shared state.
Given both a dataset and causal graph, the agent ensures consistency in variable definition, identification, and estimation strategies while explicitly reporting when analyses are not applicable.

As illustrated in Figure~\ref{fig:detail_inference_figure}, the Causal Inference Agent is organized into four modules.
(1) The \textit{Variable Selection} module grounds the user’s causal query by fixing the treatment and outcome with respect to the available data and assigns remaining variables to causal roles based on the selected graph, which users validate against the intended causal question.
(2) The \textit{Configuration Selection} module determines an appropriate identification and estimation strategy by matching data characteristics and graph structure to compatible causal assumptions.
(3) The \textit{Model Implementation} module executes effect estimation when the chosen configuration is identifiable, and explicitly reports failure when structural or statistical requirements are not met, avoiding spurious estimates.
(4) Finally, the \textit{Interpretation} module summarizes the estimated effect, identification conditions, and limitations that govern how the result should be interpreted.

ORCA enables robust causal inferences achieving consistently low absolute error(as low as $3.2\times10^{-5}$ on REEF)and maintaining high confidence interval coverage (up to 100\%).
See Appendix~\ref{appendix:causalinference_detail} and Appendix~\ref{appendix:inference-eval} for implementation details and evaluation results.

\section{User Study \& Design Implications}
\subsection{Study Design}
We conducted a user study to evaluate ORCA using a within-subjects design with twelve data-literate participants who regularly work with Python and SQL, are familiar with causal concepts, but have comparatively less hands-on experience applying causal inference methods.
Each participant completed two causal analysis tasks on the same relational database—one with ORCA and one with a baseline assistant (GPT-4o-mini) featuring tool access, step-gated progression, and minimal state persistence—in counterbalanced order.
Each task, framed as a standardized average treatment effect (ATE) query, followed a three-stage pipeline of determining the analysis dataset, specifying the causal graph, and estimating the causal effect, and was limited to 60 minutes.
We recorded interaction logs, task outcomes, and pre- and post-task survey responses.
Detailed demographics and the study protocol are provided in Appendix~\ref{appendix:user_study}.

To our knowledge, no existing benchmark provides a relational database with a known ground-truth causal graph and effect estimates, both of which are required for controlled evaluation of end-to-end causal analysis workflows.
For evaluation, we construct REEF (Relational E-commerce Evaluation Framework), a semi-synthetic relational database reflecting real-world challenges of causal analysis on relational data.
In REEF, variables relevant to a causal query are distributed across multiple tables, requiring users to determine join paths, aggregation levels, and units of analysis.
The database is generated from a predefined Structural Causal Model (SCM) that specifies causal dependencies, enabling evaluation against a known causal structure.
REEF also incorporates realistic data characteristics such as missing values and inactive entities.
See Appendix~\ref{appendix:reef} for SCM specifications and data generation procedures.

\subsection{Results and Implications}
\label{sec:user_study_results}

\begin{table}[]
    \centering
    \begin{tabular}{lcc}
    \toprule
    & \textbf{Baseline} & \textbf{ORCA}\\
    \midrule
    Task Completion Rate 
    & 0.75 & \textbf{0.83}  \\
    Discovery Success Rate
    & 0.67 & \textbf{0.83} \\
    Data Selection Success Rate 
    & 0.75 & \textbf{1.00}  \\
    CI coverage
    & 0.083 & \textbf{0.50} \\
    \bottomrule
    \end{tabular}
    \caption{The table reports task completion rate, causal discovery success, data selection success, and confidence interval (CI) coverage across conditions. 
    All metrics corresponds to better performance when they have higher values. 
    ORCA consistently outperforms the baseline across all metrics, achieving higher task completion and discovery success, perfect data selection success, and substantially improved CI coverage, indicating more reliable causal effect estimation.}
    \label{tab:user_study_subfigure}
\end{table}

\begin{table*}[t]
    \centering
    \begin{tabular}{lcccc}
    \toprule
     & \multicolumn{2}{c}{\textbf{User interaction time}}
     & \multicolumn{2}{c}{\textbf{User interaction turns}} \\
    \cmidrule(lr){2-3} \cmidrule(lr){4-5}
     & \textbf{Baseline} & \textbf{ORCA}
     & \textbf{Baseline} & \textbf{ORCA} \\
    \midrule
    Data exploration step
     & $439.34 \pm 373.32 $ & $80.60 \pm 77.71$  & $5.72 \pm 4.17$ & $1\pm2.96$  \\
    Causal discovery step
     & $78.49\pm 117.51$  & $164.02 \pm 155.64$  & $2\pm 4.14$ & $1.1\pm2.331$ \\
    Causal analysis step
     & $71.48\pm 66.69$  & $155.90 \pm 202.15$  & $1.64 \pm 1.8$  & $0.714 \pm 1.25$ \\
    \bottomrule
    \end{tabular}
    \caption{User interaction time and number of additional interaction turns across analysis stages for the baseline and ORCA conditions. 
    ORCA substantially reduces user interaction time during data exploration while maintaining comparable or fewer interaction turns across all stages.}
    \label{tab:user_interaction}
\end{table*}

\begin{figure*}
    \centering
    \includegraphics[width=\linewidth]{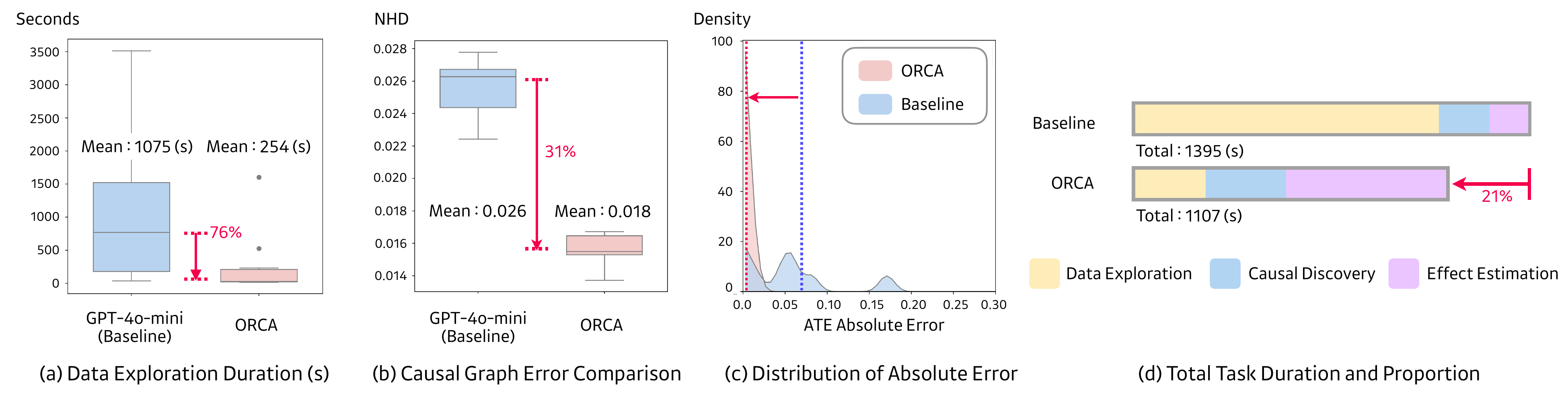}
    \caption{(a) Boxplot of time spent at data exploration analysis step. 
    ORCA significantly reduces time spent on early-stage data exploration. 
    (b) Comparison on structural error of the inferred causal graph. 
    ORCA consistently produces more accurate graphs than the baseline.
    (c) Distribution of partipants reports over Average Treatment Effect (ATE) absolute error.
    Outliers are omitted for readability.
    (d) Total task duration and proportion of total task time spent at each step. 
    ORCA shifts user effort towards human-central causal reasoning from repetitive labor-intensive data exploration.}
    \Description{This figure compares user performance between ORCA and a baseline system across time efficiency and causal analysis accuracy, using four subplots labeled (a) through (d).
Subplot (a) shows a boxplot of time spent on the data exploration stage. The baseline exhibits substantially longer and more variable exploration times, while ORCA shows consistently shorter durations, indicating a significant reduction in early-stage data exploration effort.
Subplot (b) presents a boxplot comparing causal graph error, measured as normalized Hamming distance (NHD) to the ground-truth graph. ORCA achieves lower NHD values than the baseline, indicating that the inferred causal graphs are consistently closer to the ground truth.
Subplot (c) displays the distribution of absolute error between estimated and true Average Treatment Effects (ATE). ORCA’s error distribution is concentrated near zero, while the baseline shows a wider distribution with heavier tails. Outliers are omitted for readability.
Subplot (d) shows stacked horizontal bars representing total task duration and the proportion of time spent in different stages. ORCA has a shorter total task time than the baseline and allocates a smaller proportion of time to data exploration, with relatively more time devoted to later-stage causal discovery and effect estimation.
Overall, the figure demonstrates that ORCA reduces early-stage exploration time, improves causal graph accuracy, lowers estimation error, and reallocates user effort toward analytically meaningful stages of causal reasoning.}
    \label{fig:user_study_result}
\end{figure*}

\paragraph{Findings 1: Improved causal outcome quality through reduced upstream errors.} As shown in Figure~\ref{fig:user_study_result}(c), ORCA produces ATE estimates more concentrated near zero, with a smaller mean absolute error (0.003 vs. 0.06).
Even when all trials are included, the gap becomes substantially larger (0.06 vs. 18.07), indicating that baseline errors are not only more dispersed but also prone to extreme failures.
Such failures were largely driven by cases where estimation was applied to mis-specified datasets or under unverified assumptions that violated identification conditions, a failure mode ORCA is specifically designed to prevent.
Consistent with this, ORCA produces causal graphs with 31\% lower Normalized Hamming Distance (see Figure~\ref{fig:user_study_result}(b)), indicating that structural improvements earlier in the pipeline carry forward into more reliable estimates.
As shown in Table~\ref{tab:user_study_subfigure}, 50\% of ORCA’s ATE estimates successfully fall within the confidence interval of the ground-truth effect, compared to 8\% for the baseline.
These improvements reflect cumulative gains from upstream stages, ORCA mitigating error propagation via step-wise pipeline that prevents users from skipping analytically required steps.
The system prevents users from carrying forward unchecked assumptions and ensures consistency across stages, thereby directly shaping downstream feasibility and correctness.

\paragraph{Findings 2: Reallocation of user effort and reduced trial-and-error through structured interaction.} ORCA reshapes how users interact with the causal analysis workflow by making early decisions explicit and surfacing errors earlier.
Under the baseline condition, users spent up to 77\% of their total analysis time to repeated data extraction (see Figures~\ref{fig:user_study_result}(a, d)), often leading to backtracking due to omitted variables or overly small subsets.
In contrast, ORCA reduces exploration time by approximately 76\% by allowing users to reallocate effort toward more critical stages such as assumption validation and result interpretation.
This shift is accompanied by fewer corrective revisions in later stages.
Interaction logs show that baseline users issued revision requests on average of 2.0 times during causal discovery and 1.64 times during effect estimation, whereas ORCA users required fewer revisions, with averages of 1.1 and 0.71, respectively.
Table~\ref{tab:user_interaction} shows that, instead of repeatedly revising outputs, ORCA users spent modestly more time inspecting intermediate artifacts to confirm that each step had been completed as intended.
Even with the more time spent in causal reasoning, total time spent on analysis was reduced by 21\%.
Survey responses corroborate these findings, with participants rating ORCA substantially higher in transparency (5.22 vs. 3.28) and trust (5.13 vs. 3.33), alongside reduced trial-and-error compared to the baseline (see Tables~\ref{tab:appendix_post_survey},~\ref{tab:appendix_post_comparison}

\section{Conclusion and Future Work}
We present ORCA, an interactive multi-agent framework for end-to-end causal analysis on relational databases.
Motivated by limitations in existing LLM agent workflows, ORCA organizes analysis as a dependency-aware, step-wise workflow with shared state and human-verified checkpoints, preventing early specification errors from propagating downstream.
Evaluations and a controlled user study show that this orchestration improves both efficiency and causal outcome quality, yielding lower ATE error and higher task completion rates.
To support evaluation in realistic relational settings, we introduce REEF, a semi synthetic relational e-commerce benchmark with an explicit structural causal model.
Future work includes exploring more flexible iteration across stages while preserving verifiability, evaluating ORCA with larger and more diverse user populations, and extending the framework to additional tasks and domains.

\begin{acks}
This work was supported by Institute of Information \& communications Technology Planning \& Evaluation(IITP) grant funded by the Korea government(MSIT) (No\. RS-2022-0-00612)
This work was supported by the National Research Foundation of Korea(NRF) grant funded by the Korea government(MSIT) (No\. RS-2024-00410082)
This work was supported by the National Research Foundation of Korea (NRF) grant funded by the Korea government (MSIT)(No\. NRF-2022M3J6A1063595)
\end{acks}

\bibliographystyle{ACM-Reference-Format}
\bibliography{reference}

@String{Computing = "Computing" }

@String{Springer = "Springer-Verlag" }

@inproceedings{wu2024autogen,
  author    = {Wu, Qingyun and Bansal, Gagan and Zhang, Jieyu and Wu, Yiran and Li, Beibin and Zhu, Erkang and Jiang, Li and Zhang, Xiaoyun and Zhang, Shaokun and Liu, Jiale},
  title     = {AutoGen: Enabling Next-Gen {LLM} Applications via Multi-Agent Conversations},
  booktitle = {Proceedings of the First Conference on Language Modeling},
  year      = {2024}
}

@misc{sharma2020dowhy,
  author = {Sharma, Amit and Kiciman, Emre},
  title  = {{DoWhy}: An End-to-End Library for Causal Inference},
  year   = {2020},
  eprint = {2011.04216},
  archivePrefix = {arXiv}
}

@misc{Li2023CanLA,
  author = {Li, Jinyang and Hui, Binyuan and Qu, Ge and Li, Binhua and Yang, Jiaxi and Li, Bowen and Wang, Bailin and Qin, Bowen and Cao, Rongyu and Geng, Ruiying and Huo, Nan and Ma, Chenhao and Chang, Kevin C. and Huang, Fei and Cheng, Reynold and Li, Yongbin},
  title  = {Can {LLM}s Already Serve as a Database Interface? A Big Bench for Large-Scale Database Grounded Text-to-{SQL}},
  year   = {2023},
  eprint = {2305.03111},
  archivePrefix = {arXiv}
}

@misc{guo2024large,
  author = {Guo, Taicheng and Chen, Xiuying and Wang, Yaqi and Chang, Ruidi and Pei, Shichao and Chawla, Nitesh V. and Wiest, Olaf and Zhang, Xiangliang},
  title  = {Large Language Model Based Multi-Agents: A Survey of Progress and Challenges},
  year   = {2024},
  eprint = {2402.01680},
  archivePrefix = {arXiv}
}

@misc{langgraph2024,
  author = {LangChain},
  title  = {LangGraph Studio},
  year   = {2024},
  url    = {https://github.com/langchain-ai/langgraph}
}

@misc{dong2022high,
  author = {Dong, Shuyu and Uemura, Kento and Fujii, Akito and Chang, Shuang and Koyanagi, Yusuke and Maruhashi, Koji and Sebag, Mich{\`e}le},
  title  = {High-Dimensional Causal Discovery: Learning from Inverse Covariance via Independence-Based Decomposition},
  year   = {2022},
  eprint = {2211.14221},
  archivePrefix = {arXiv}
}

@article{berkessa2024review,
  author  = {Berkessa, Zewude A. and L{\"a}{\"a}r{\"a}, Esa and Waldmann, Patrik},
  title   = {A Review of Causal Methods for High-Dimensional Data},
  journal = {IEEE Access},
  year    = {2024},
  publisher = {IEEE}
}

@article{mondal2025fast,
  author  = {Mondal, Sakib A. and Rv, Prashanth and Rao, Sagar},
  title   = {A Fast Algorithm for High-Dimensional Causal Discovery},
  journal = {International Journal of Data Science and Analytics},
  volume = {20},
  pages   = {1--10},
  year    = {2025}
}

@article{maojun2025lambda,
  author  = {Sun, Maojun and Han, Ruijian and Jiang, Binyan and Qi, Houduo and Sun, Defeng and Yuan, Yancheng and Huang, Jian},
  title   = {Lambda: A Large Model Based Data Agent},
  journal = {Journal of the American Statistical Association},
  pages   = {1--20},
  year    = {2025}
}

@misc{testini2025measuring,
  author = {Testini, Irene and Hern{\'a}ndez-Orallo, Jos{\'e} and Pacchiardi, Lorenzo},
  title  = {Measuring Data Science Automation: A Survey of Evaluation Tools for {AI} Assistants and Agents},
  year   = {2025},
  eprint = {2506.08800},
  archivePrefix = {arXiv}
}

@misc{kulkarni2025agent,
  author = {Kulkarni, Mandar},
  title  = {Agent-s: {LLM} Agentic Workflow to Automate Standard Operating Procedures},
  year   = {2025},
  eprint = {2503.15520},
  archivePrefix = {arXiv}
}

@article{kuhne2022causal,
  author  = {K{\"u}hne, F. and Schomaker, M. and Stojkov, I. and Jahn, B. and Conrads-Frank, A. and Siebert, S. and Sroczynski, G. and Puntscher, S. and Schmid, D. and Schnell-Inderst, P. and Siebert, U.},
  title   = {Causal Evidence in Health Decision Making},
  journal = {German Medical Science},
  volume  = {20},
  pages   = {Doc12},
  year    = {2022},
  doi     = {10.3205/000314}
}

@article{bojinov2020importance,
  author  = {Bojinov, Iavor I. and Chen, Albert and Liu, Min},
  title   = {The Importance of Being Causal},
  journal = {Harvard Data Science Review},
  volume  = {2},
  number  = {3},
  year    = {2020},
  publisher = {The MIT Press}
}

@inproceedings{wu2022ai,
  title={Ai chains: Transparent and controllable human-ai interaction by chaining large language model prompts},
  author={Wu, Tongshuang and Terry, Michael and Cai, Carrie Jun},
  booktitle={Proceedings of the 2022 CHI conference on human factors in computing systems},
  pages={1--22},
  year={2022}
}

@inproceedings{epperson2025interactive,
  title={Interactive debugging and steering of multi-agent ai systems},
  author={Epperson, Will and Bansal, Gagan and Dibia, Victor C and Fourney, Adam and Gerrits, Jack and Zhu, Erkang and Amershi, Saleema},
  booktitle={Proceedings of the 2025 CHI Conference on Human Factors in Computing Systems},
  pages={1--15},
  year={2025}
}

@article{heckman2022econometric,
  author  = {Heckman, James J. and Pinto, Rodrigo},
  title   = {The Econometric Model for Causal Policy Analysis},
  journal = {Annual Review of Economics},
  volume  = {14},
  number  = {1},
  pages   = {893--923},
  year    = {2022}
}

@inproceedings{chan2024autocd,
  author    = {Chan, Gerlise and Claassen, Tom and Hoos, Holger and Heskes, Tom and Baratchi, Mitra},
  title     = {{AutoCD}: Automated Machine Learning for Causal Discovery Algorithms},
  booktitle = {AutoML 2024 Methods Track},
  year      = {2024}
}

@misc{han2024causal,
  author = {Han, Kairong and Kuang, Kun and Zhao, Ziyu and Ye, Junjian and Wu, Fei},
  title  = {Causal Agent Based on Large Language Models},
  year   = {2024},
  eprint = {2408.06849},
  archivePrefix = {arXiv}
}

@misc{le2024multi,
  author = {Le, Hao Duong and Xia, Xin and Chen, Zhang},
  title  = {Multi-Agent Causal Discovery Using Large Language Models},
  year   = {2024},
  eprint = {2407.15073},
  archivePrefix = {arXiv}
}

@book{peters2017elements,
  title={Elements of causal inference: foundations and learning algorithms},
  author={Peters, Jonas and Janzing, Dominik and Sch{\"o}lkopf, Bernhard},
  year={2017},
  publisher={The MIT press}
}

@article{jin2023can,
  title={Can large language models infer causation from correlation?},
  author={Jin, Zhijing and Liu, Jiarui and Lyu, Zhiheng and Poff, Spencer and Sachan, Mrinmaya and Mihalcea, Rada and Diab, Mona and Sch{\"o}lkopf, Bernhard},
  journal={arXiv preprint arXiv:2306.05836},
  year={2023}
}

@article{khatibi2024alcm,
  title={Alcm: Autonomous llm-augmented causal discovery framework},
  author={Khatibi, Elahe and Abbasian, Mahyar and Yang, Zhongqi and Azimi, Iman and Rahmani, Amir M},
  journal={arXiv preprint arXiv:2405.01744},
  year={2024}
}

@article{shimizu2006linear,
  title={A linear non-Gaussian acyclic model for causal discovery.},
  author={Shimizu, Shohei and Hoyer, Patrik O and Hyv{\"a}rinen, Aapo and Kerminen, Antti and Jordan, Michael},
  journal={Journal of Machine Learning Research},
  volume={7},
  number={10},
  year={2006}
}

@article{zhang2012identifiability,
  title={On the identifiability of the post-nonlinear causal model},
  author={Zhang, Kun and Hyvarinen, Aapo},
  journal={arXiv preprint arXiv:1205.2599},
  year={2012}
}

@article{tsamardinos2006max,
  title={The max-min hill-climbing Bayesian network structure learning algorithm},
  author={Tsamardinos, Ioannis and Brown, Laura E and Aliferis, Constantin F},
  journal={Machine learning},
  volume={65},
  number={1},
  pages={31--78},
  year={2006},
  publisher={Springer}
}

@article{zheng2024causal,
  title={Causal-learn: Causal discovery in python},
  author={Zheng, Yujia and Huang, Biwei and Chen, Wei and Ramsey, Joseph and Gong, Mingming and Cai, Ruichu and Shimizu, Shohei and Spirtes, Peter and Zhang, Kun},
  journal={Journal of Machine Learning Research},
  volume={25},
  number={60},
  pages={1--8},
  year={2024}
}

@article{bach2022doubleml,
  title={DoubleML-an object-oriented implementation of double machine learning in python},
  author={Bach, Philipp and Chernozhukov, Victor and Kurz, Malte S and Spindler, Martin},
  journal={Journal of Machine Learning Research},
  volume={23},
  number={53},
  pages={1--6},
  year={2022}
}

@article{wager2018estimation,
  title={Estimation and inference of heterogeneous treatment effects using random forests},
  author={Wager, Stefan and Athey, Susan},
  journal={Journal of the American Statistical Association},
  volume={113},
  number={523},
  pages={1228--1242},
  year={2018},
  publisher={Taylor \& Francis}
}

@article{kunzel2019metalearners,
  title={Metalearners for estimating heterogeneous treatment effects using machine learning},
  author={K{\"u}nzel, S{\"o}ren R and Sekhon, Jasjeet S and Bickel, Peter J and Yu, Bin},
  journal={Proceedings of the national academy of sciences},
  volume={116},
  number={10},
  pages={4156--4165},
  year={2019},
  publisher={National Academy of Sciences}
}

@article{peters2015structural,
  title={Structural intervention distance for evaluating causal graphs},
  author={Peters, Jonas and B{\"u}hlmann, Peter},
  journal={Neural computation},
  volume={27},
  number={3},
  pages={771--799},
  year={2015},
  publisher={MIT Press}
}

@article{zhou2025guardian,
  title={GUARDIAN: Safeguarding LLM Multi-Agent Collaborations with Temporal Graph Modeling},
  author={Zhou, Jialong and Wang, Lichao and Yang, Xiao},
  journal={arXiv preprint arXiv:2505.19234},
  year={2025}
}

@inproceedings{salimi2020causal,
  title={Causal relational learning},
  author={Salimi, Babak and Parikh, Harsh and Kayali, Moe and Getoor, Lise and Roy, Sudeepa and Suciu, Dan},
  booktitle={Proceedings of the 2020 ACM SIGMOD international conference on management of data},
  pages={241--256},
  year={2020}
}

@article{holland1986statistics,
  title={Statistics and causal inference},
  author={Holland, Paul W},
  journal={Journal of the American statistical Association},
  volume={81},
  number={396},
  pages={945--960},
  year={1986},
  publisher={Taylor \& Francis}
}

@inproceedings{ahsan2023learning,
  title={Learning relational causal models with cycles through relational acyclification},
  author={Ahsan, Ragib and Arbour, David and Zheleva, Elena},
  booktitle={Proceedings of the AAAI Conference on Artificial Intelligence},
  volume={37},
  number={10},
  pages={12164--12171},
  year={2023}
}

@article{negro2025relational,
  title={Relational Causal Discovery with Latent Confounders},
  author={Negro, Matteo and Piras, Andrea and Ahsan, Ragib and Arbour, David and Zheleva, Elena},
  journal={arXiv preprint arXiv:2507.01700},
  year={2025}
}

@article{maier2013sound,
  title={A sound and complete algorithm for learning causal models from relational data},
  author={Maier, Marc and Marazopoulou, Katerina and Arbour, David and Jensen, David},
  journal={arXiv preprint arXiv:1309.6843},
  year={2013}
}

\appendix
\newpage

\renewcommand{\thefigure}{\Alph{section}.\arabic{figure}}
\renewcommand{\thetable}{\Alph{section}.\arabic{table}}

\section{Related Works}

\subsection{Multi-Agentic system and HITL} 
The rapid advancement of large-scale language models (LLMs) has shifted their use from single-model response generation to multi-agent systems that integrate tool usage and stateful interaction.
These systems are increasingly realized as multi-agent frameworks, where multiple specialized agents collaborate to decompose and solve complex tasks.
Representative frameworks such as AutoGen~\cite{wu2024autogen} and LangGraph~\cite{langgraph2024} exemplify this paradigm by enabling structured interaction among agents with distinct roles.
Despite these advancements, multi-agent systems inherit fundamental limitations of their underlying LLMs. 
Problems such as hallucination, and unstable reasoning are often amplified through cascading interactions, leading to error propagation across the workflow~\cite{zhou2025guardian,guo2024large}.
As multi-agent systems complexity grows, mitigating such error accumulation has become a central research concern.

Moving beyond fully autonomous paradigms, emerging studies underscore the necessity of integrating human–AI interaction within multi-agent workflows~\cite{epperson2025interactive}. 
Rather than passively observing agent outcomes, this new direction emphasizes interactive mechanisms that enable users to inspect system internal states, correcting erroneous trajectories, and dynamically steering analytical processes.
These Human-in-the-Loop (HIL) mechanisms provide a foundation for more transparent and controllable multi-agent systems~\cite{wu2022ai}.
Core functionalities include pausing execution, rewinding to previous states, and editing prior messages to alter workflow trajectories, enabling localization of failure points and real-time verification of corrective interventions~\cite{epperson2025interactive}.

\subsection{Automating Causal methods}
Causal analysis plays a critical role in reliable decision-making by uncovering the underlying cause-effect mechanisms beyond mere correlations, and is widely recognized as essential for improving model interpretability across domains.
Despite its importance, causal discovery and inference remain fundamentally challenging in real-world settings.
Observational data alone makes it difficult to control for latent confounders and nonlinear dependencies, while the combinatorial explosion of potential directed acyclic graphs (DAGs) further increases computational complexity. 
Moreover, the performance of causal discovery algorithms varies considerably depending on data characteristics (continuous, discrete, or mixed types), and the process of method selection and hyperparameter tuning also requires substantial domain expertise. 
Consequently, recent research has increasingly explored LLM-based approaches and task-level automation to reduce expert burden in causal analysis (e.g. dowhy~\cite{sharma2020dowhy}, causallearn~\cite{zheng2024causal}).

Traditional causal discovery methods include constraint-based, score-based, and functional causal model (FCM) approaches, but their performance varies widely across datasets and often requires manual configuration.
To address these limitations, prior work on automated causal discovery has focused on reducing manual model selection and configuration.
AutoCD~\cite{chan2024autocd} automated the selection and tuning of causal discovery algorithms via Bayesian Optimization, addressing performance variability across datasets.
Parallel to these efforts, the integration of LLMs into causal reasoning workflows has attracted growing attention. 
The Causal Agent~\cite{han2024causal} leverages LLMs to automate causal analysis on tabular data using external tools, demonstrating that LLMs can effectively decompose and analyze causal problems.
Also, Multi-Agent Causal Discovery Framework (MAC)~\cite{le2024multi} employs multiple LLM agents to collaboratively generate and validate causal graphs through structured debate.

Collectively, these studies signify the emergence of an Agentic System paradigm for partially automating causal analysis tasks.
This direction lays the foundation for next-generation causal reasoning systems that combine automation, adaptability, and human interpretability within a unified framework.

\subsection{Causal Inference over Relational Databases}
Causal inference over relational databases has become increasingly important as real-world data are stored across multiple interconnected tables; however, traditional causal analysis tools struggle to handle the complexity of such multi-relational structures~\cite{salimi2020causal}.
Standard causal discovery methods typically assume a single i.i.d. dataset~\cite{holland1986statistics}, making them ill-suited for relational settings where observations are not independent across tables.

To bridge this gap, recent work has proposed relational causal discovery algorithms that extend classical techniques to multi-table data.
For example, ~\citet{ahsan2023learning} enable causal discovery in cyclic relational models by characterizing conditions under which RCD~\cite{maier2013sound} remains sound despite mutual dependencies.
In addition, ~\citet{negro2025relational} propose a relational FCI method (RelFCI) to account for latent confounders that span multiple linked tables.

In practice, causal analysis over relational databases further requires validation of modeling assumptions which are often implicit and error-prone in real-world data.
However, even with existing advances in relational causal discovery, research on full workflow supporting causal analysis directly over relational databases remains an open and active area.

\section{Implementation Details of ORCA framework}
\setcounter{figure}{0}
\setcounter{table}{0}
\label{appendix:framework_detail}
This section provides implementation-level details of the ORCA framework to support reproducibility. 
We describe the concrete inputs, outputs, prompts, diagnostics, and control logic used by each agent, complementing the high-level system description presented in Section~\ref{sec:method}. 
Unless otherwise noted, this section focuses on how each component is instantiated in practice rather than on design motivations, which are discussed in the main text.
To support transparency and reproducibility, we release the full implementation of ORCA, including all agents and orchestration logic.\footnote{\url{https://anonymous.4open.science/r/ORCA-E70F}}

\subsection{Details for the Data Exploration Agent}
\label{appendix:dataexploration_detail}

\subsubsection{Table Explorer} 
Given a user-specified table, \textit{Table Explorer} operates over database metadata, including schema definitions, column-level statistics, and foreign key relationships.
Using the prompt shown in Prompt~\ref{fig:table_explorer_prompt}, it summarizes table- and column-level properties such as null ratios, distinct counts, and distributional characteristics.
These summaries surface constraints on how columns can be used in analysis.
For example, columns with near-zero variance are flagged as invalid candidates for treatments or outcomes, while columns with high missingness flags potential data quality concerns.
The module further examines the table’s relational context by identifying adjacent tables.
This step allows users to anticipate how joins may change the unit of analysis or introduce duplication.
Finally, the \textit{Table Explorer} synthesizes statistical and relational signals into an explicit assessment of data suitability, highlighting integrity issues and structural risks relevant to subsequent steps.
An example of the output is shown in Figure~\ref{fig:table-explorer-log}.

\subsubsection{Table Recommender} 
\textit{Table Recommender} takes as input a natural language query or analysis plan, together with database metadata which includes schema information and foreign key relationships, and diagnostics from the \textit{Table Explorer}.
Guided by Prompt~\ref{fig:table_recommender_prompt}, it first extracts an analytical objective by identifying target entities, outcomes of interest, and required information types (e.g., behavioral, transactional, or temporal).
It then narrows the search space via semantic matching over table and column metadata to identify candidate tables relevant to the analysis and filter out unrelated tables.
Finally, the module explores foreign key paths among candidate tables to assess relational connectivity. 
This analysis verifies that necessary join paths exist and makes explicit how tables are connected. 
As a result, users can identify relational paths that may introduce confounding, as well as tables that may lead to unnecessary joins.
Based on these steps, the \textit{Table Recommender} produces a ranked list of tables and an ERD visualizing their connections, helping users pinpoint database components essential to their research goals and validate the relational scope before downstream analysis.
See Figure~\ref{fig:table-rec-log} for detailed output.

\subsubsection{Text2SQL Generator} 
\textit{Text2SQL Generator} takes as input the user query, the set of tables selected by the \textit{Table Recommender}, and database metadata.
Guided by the generation prompt shown in Figure~\ref{fig:text2sql_generation_prompt}, it first parses the natural language request into a structured query specification, identifying target entities, variables of interest, temporal constraints, and the intended unit of analysis.
Based on this specification, it builds a query skeleton by determining appropriate \texttt{SELECT}, \texttt{JOIN}, \texttt{WHERE}, and \texttt{GROUP BY} clauses, while restricting joins to the recommended tables and valid schema relationships.
To ensure robustness, the module incorporates an automated execution-feedback loop, which iteratively corrects errors in column references, join conditions, or aggregation logic via a self-correction prompt (Figure~\ref{fig:text2sql_selfcorrection_prompt}).
The module further validates whether the executed result aligns with the original analytical intent using an intent-validation prompt (Figure~\ref{fig:text2sql_validation_prompt}), preventing cases where a query is technically valid but semantically misaligned.
This structured and automated refinement reduces the need for manual trial-and-error.
The Text2SQL Generator ultimately returns an executable SQL query along with its resulting dataset. 
This output provides a well-defined and verifiable dataset specification for subsequent causal discovery and causal inference.
An example execution trace and generated output are shown in Figure~\ref{fig:text2sql-generator-log}.

\subsubsection{Data Preprocessor} 
\textit{Data Preprocessor} applies data transformation including outlier handling, normalization, and format alignment to produce clean, analysis-ready datasets.
As part of this process, it performs schema detection to annotate variable information.
Specifically, the module infers data types and computes categorical cardinalities, providing the specifications required for Causal Discovery.
This module serves as a final checkpoint before modeling, ensuring that analytical assumptions align with the prepared dataset.

\subsection{Details for the Causal Discovery Agent}
\label{appendix:causaldiscovery_detail}
\subsubsection{Testing Causal Hypothesis}
The first step performs a comprehensive statistical diagnosis to assess the validity of key causal assumptions underlying different discovery algorithms.
To this end, the module is organized as a series of diagnostic tests that evaluate data properties relevant to these assumptions.
This step produces a structured data profile that summarizes assumption-relevant properties of the dataset through three hierarchical checks:
\begin{enumerate}
\item \textbf{Basic Checks.} 
The dataset is first classified as \textit{Pure Continuous}, \textit{Pure Categorical}, or \textit{Mixed}.
Then the reliability of Conditional Independence (CI) tests is assessed by comparing the sample size ($n$) to the number of variables ($p$); when $p > n$, CI reliability is flagged critically low, prompting conservative algorithm choices.
High-cardinality categorical variables are also detected to avoid computational bottlenecks in discrete or mixed-type algorithms.

\item \textbf{Global Tests.} 
Dataset-level assumptions are evaluated using multivariate tests. 
For continuous data, the Henze–Zirkler (HZ) test is applied to assess multivariate normality, where $p < 0.05$ indicates deviation from Gaussianity.
To evaluate global linearity, the Ramsey RESET test is performed on continuous variables and one-hot encoded mixed variables.

\item \textbf{Pairwise Profiles:}
This check analyzes all pairs of continuous variables.
\begin{itemize}
    \item \textbf{Linearity Score ($S_{lin}$)} 
    For each variable pair, linearity is tested via a nested comparison between a linear model and a spline-based Generalized Additive Model (GAM).
    Let $p_{\text{nonlin}}$ denote the p-value for the smooth term; a linearity indicator is defined as $\mathbb{I}_{\text{lin}} = \mathbf{1}(p_{\text{nonlin}} > \alpha)$ with $\alpha = 0.05$.

    \item \textbf{Non-Gaussianity Score ($S_{ng}$)}
    For each pair, we test residual normality using the Jarque–Bera and Shapiro–Wilk tests.
    Let $p_{\text{val}}$ be the resulting p-value and define $S_{\text{ng}} = 1 - p_{\text{val}}$, where larger values indicate stronger evidence of non-Gaussian noise.
    
    \item \textbf{ANM Score:} 
    Consistency with the Additive Noise Model is assessed by testing independence between predictors and residuals using HSIC.
    The resulting p-value is used as $S_{\text{anm}}$, with higher values indicating stronger support for the ANM assumption.
\end{itemize}
\end{enumerate}
For the linearity score, pairwise binary indicators are aggregated by their mean, yielding the proportion of variable pairs consistent with linearity.
For non-Gaussianity and ANM scores, pairwise values are aggregated using the median to ensure robustness against unstable or weakly related variable pairs.

\subsubsection{Algorithm Configuration}
\begin{figure*}[t]
    \centering
    \includegraphics[width=\linewidth]{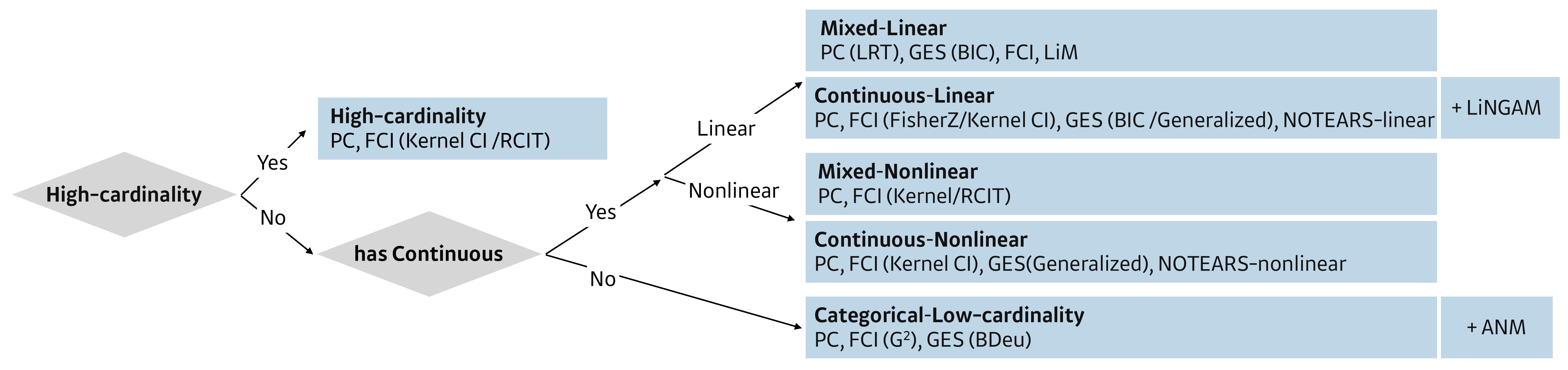}
    \caption{Decision-tree of ORCA’s algorithm configuration.
    The tree selects a base causal discovery scenario based on categorical cardinality, the data type(presence of continuous variables), and linearity diagnostics.
    Leaf nodes summarize compatible algorithm families, with additional assumption checks to conditionally add optional methods (LiNGAM and ANM).}
    \Description{A left-to-right decision tree illustrating how ORCA configures causal discovery algorithms.
    The tree begins with a binary decision on whether the dataset contains high-cardinality categorical variables.
    If yes, the configuration selects high-cardinality methods such as PC or FCI with kernel-based conditional independence tests.
    If no, the tree next checks whether the dataset includes continuous variables.
    For datasets with continuous variables, an additional split distinguishes between linear and nonlinear relationships.
    Linear continuous data lead to mixed-linear or continuous-linear configurations, including PC or FCI with linear tests, GES, and optionally LiNGAM when its assumptions are satisfied.
    Nonlinear continuous data lead to mixed-nonlinear or continuous-nonlinear configurations, including PC or FCI with kernel tests, GES, and nonlinear NOTEARS.
    If the dataset does not contain continuous variables, the configuration selects categorical low-cardinality methods such as PC or FCI with G-squared tests and GES with BDeu scoring, with ANM optionally added under suitable assumptions.
    Leaf nodes list the corresponding families of causal discovery algorithms for each scenario.}
    \label{fig:alg_config_figure}
\end{figure*}

Based on the data profile, the \textit{Algorithm Configuration} module determines a tailored execution plan.
This step maps statistical properties to a portfolio of compatible algorithms, ensuring that each solver is applied only when its underlying assumptions are satisfied.
The agent evaluates properties such as data type, cardinality, and the aggregated linearity and Gaussianity scores to determine the appropriate configuration, following the branching logic illustrated in Figure~\ref{fig:alg_config_figure}.
Thresholds on aggregated diagnostic scores are applied as implementation-level criteria.
Specifically, we use a fixed threshold $\tau_{lin} = 0.5$ on the aggregated linearity score, corresponding to a majority of variable pairs exhibiting linear behavior.
Similarly, $\tau_{ng} = \tau_{anm} = 0.5$ are used as majority-based thresholds for non-Gaussianity and additive noise diagnostics, which guide the conditional inclusion of additional discovery methods.
These values were fixed by selecting the setting that achieved the lowest and most stable SHD/SID among candidate thresholds.

\subsubsection{Graph Generation and Scoring}
The third step generates candidate causal graphs according to the execution plan and evaluates their plausibility using stability and consistency based criteria.
The execution module runs the algorithms specified in the plan in parallel on the preprocessed data.
The output is a set of candidate graphs $G=\{G_{1}, G_{2}, \dots, G_{k}\}$, which may include DAGs, CPDAGs, or PAGs.
Because different algorithms rely on distinct optimization criteria and scoring functions, their outputs are not directly comparable.
Instead, ORCA evaluates each candidate graph based on its agreement with the observed data distribution and its stability under perturbations, using a composite score derived from three metrics: 
\begin{enumerate}
\item \textbf{Markov Consistency ($S_{\text{mc}}$):}
This metric measures whether the conditional independence relations implied by a graph are supported by the data.
For a given graph $G$, we extract a local Markov basis of implied d-separation statements and test each statement.
The score is defined as
\begin{equation}
S_{\text{mc}}(G)
=
1 -
\frac{N_{\text{rejected}}}{N_{\text{tested}}},
\end{equation}
where $N_{\text{rejected}}$ denotes the number of conditional independence hypotheses rejected by the data at significance level $\alpha$, and $N_{\text{tested}}$ is the total number of tested statements.
For graph types that do not encode a full set of testable conditional independencies (e.g., PAGs), this metric is omitted.

\item \textbf{Sampling Stability ($S_{\text{samp}}$):}  
To assess robustness to sampling variability, we apply bootstrap resampling.
Given $B$ bootstrap samples, the same discovery algorithm is re-run to obtain an ensemble $\{G^{*}_{1}, \dots, G^{*}_{B}\}$.
The confidence of an edge $e$ is defined as its empirical frequency $P(e) = \frac{1}{B} \sum_{b=1}^{B} \mathbb{I}\!\left(e \in G^{*}_{b}\right).$
The sampling stability score of graph $G$ is the average confidence over its edges:
\begin{equation}
    S_{\text{samp}}(G)
    =
    \frac{1}{|E_{G}|}
    \sum_{e \in E_{G}} P(e).
\end{equation}

\item \textbf{Structural Stability ($S_{\text{struct}}$):}  
This metric evaluates sensitivity to variable selection.
We generate random subsets of variables $V_i \subset V$, re-run the algorithm on each subset to obtain $G_i$, and compare it with the restriction of the original graph $G|_{V_i}$.
Instability is quantified using a graph-type–appropriate Structural Hamming Distance (SHD)~\cite{tsamardinos2006max}.
For partially oriented graphs (e.g., CPDAGs), SHD is computed on adjacencies only.

\end{enumerate}

The final \textbf{Composite Score} is computed as a weighted sum of the applicable metrics:
\begin{equation}
    \text{Score}(G)
    =
    w_{1} S_{\text{mc}} 
    + w_{2} S_{\text{samp}} 
    + w_{3} S_{\text{struct}}
\end{equation}
where the weights are normalized over the metrics applicable to the graph type.
Candidate graphs are ranked by this score, and the top-$k$ graphs are forwarded to the synthesis step.
In our implementation, we use fixed weights $(w_1, w_2, w_3) = (0.4, 0.3, 0.3)$.
These weights balance consistency with the observed data and stability under perturbations, and are kept fixed across all experiments.

\subsubsection{Ensemble and Synthesis}
The final step integrates multiple candidate causal graphs into a unified representation by aggregating structural evidence and resolving directional conflicts.

\paragraph{Consensus PAG Construction.} ORCA first builds a consensus skeleton by identifing robust adjacencies shared across candidate graphs.
For each variable pair $(X, Y)$, an undirected confidence score is computed by aggregating the normalized composite scores of candidate graphs that contain the corresponding adjacency.
If the aggregated confidence exceeds a predefined threshold $\tau$ (e.g., $\tau = 0.5$), the adjacency is retained.
Given the retained adjacencies, ORCA derives a PAG-like intermediate graph.
For each adjacency $(X, Y)$, directional evidence is aggregated using weighted voting over the top-$k$ candidate graphs.
Let $w_i$ denote the normalized composite score of graph $G_i$.
Only explicitly directed edges ($\rightarrow$) contribute to the vote.
If weighted support for one direction exceeds that of the opposite direction, the edge is oriented accordingly; otherwise, it is marked as uncertain (e.g., $X \circ\!-\!\circ Y$).
The resulting \texttt{consensus\_pag} preserves both robust structural relations and directional uncertainty where the ensemble does not provide sufficient agreement.

\paragraph{Actionable DAG Synthesis.} To enable downstream causal inference, ORCA converts the consensus PAG into a single directed acyclic graph (DAG).
Edges are processed in descending order of confidence and added incrementally under an explicit acyclicity constraint.
Edges with resolved directions are adopted whenever they do not introduce cycles.
For uncertain or conflicting edges, ORCA applies a deterministic, rank-consistent tie-breaking strategy: the direction proposed by the highest-ranked candidate graph that specifies an explicit orientation is selected.
If this choice introduces a cycle, the opposite orientation is attempted; if both violate acyclicity, the edge is discarded.
If no candidate provides directional evidence, ORCA attempts a cycle-safe orientation; otherwise, the edge is dropped.
The resulting graph is guaranteed to be fully directed and acyclic.

As a result, the \textbf{Causal Discovery Agent} outputs two complementary artifacts:
a \texttt{consensus\_pag} that preserves structural uncertainty for inspection, and a fully directed \texttt{final\_dag} used for downstream causal effect estimation.
An example execution log is shown in Figure~\ref{fig:discovery-log}.

\subsection{Details for the Causal Inference Agent}
\label{appendix:causalinference_detail}

\subsubsection{Variable Selection}
The goal of the \textit{Variable Selection} module is to parse the user’s question as a causal query by fixing the meaning of the treatment(T) and outcome(Y) with respect to the available data.
This step ensures that subsequent identification and estimation strategies operate over well-defined variables and reduces failures caused by mismatches between user intent and data representation.
Given the user query, the module utilizes Prompt~\ref{fig:variable_selection_prompt} to select treatment and outcome variables using the available columns, a schema summary, and a sample of the data.
Using the selected causal graph, remaining variables are then assigned causal roles according to their structural relationships to $T$ and $Y$, including confounders (common causes of $T$ and $Y$), mediators (post-treatment variables on causal paths from $T$ to $Y$), valid instruments, and variables that should not be conditioned on, such as colliders or descendants of the outcome.

\subsubsection{Configuration Selection}
The \textit{Configuration Selection} module selects one of several predefined tasks, employing Prompt~\ref{fig:config_selector_prompt} that lists valid options with concise descriptions. 
Then, the identification strategy is selected from backdoor, frontdoor, instrumental variable, mediation, or the ID algorithm.
Candidate confounders, mediators, and instruments identified in the previous stage inform this choice.
For example, when no valid instrumental variables are present, IV-based strategies are excluded.
Estimator selection is guided by the outcome type and data characteristics; for instance, binary outcomes favor generalized linear models, while continuous outcomes favor linear regression-based estimators.

\subsubsection{Model Implementation}
The \textit{Model Implementation} module first converts the selected causal graph into a DoWhy-compatible representation.
It then checks basic structural requirements implied by the chosen identification strategy, such as the existence of a directed path from the treatment to the outcome or the validity of candidate instruments.
If these requirements are not satisfied, the module records and reports the reason for failure.
When the requirements are satisfied, the module constructs a causal model, identifies the estimand, and further applies the selected estimator and refutation when specified.
If refutation methods are specified, DoWhy refuters are executed and their outcomes are summarized.
The resulting model object, estimand, and estimate with confidence intervals are stored. 
If estimation fails, the failure and its cause are explicitly reported.
Optional fallback results (e.g., regression-based adjustments) are provided only for descriptive references and are distinguished from causal estimates.

\subsubsection{Interpretation}
\textit{Interpretation} module uses an LLM with Prompt~\ref{fig:interpretation_prompt} to generate an understandable explanation of causal inference results based on the outputs of previous stages. 
Rather than merely verbalizing numerical estimates, it summarizes the assumptions, identification conditions, and limitations that govern how the result should be interpreted.
By making these conditions explicit, the module helps users assess the credibility and scope of the causal claim.
A sample output is provided in log~\ref{fig:analyzer-log}.

\section{REEF}
\label{appendix:reef}
\setcounter{figure}{0}
\setcounter{table}{0}
\subsection{Database Schema and Relational Structure}

\begin{figure*}[t]
    \centering
    \includegraphics[width=0.9\linewidth]{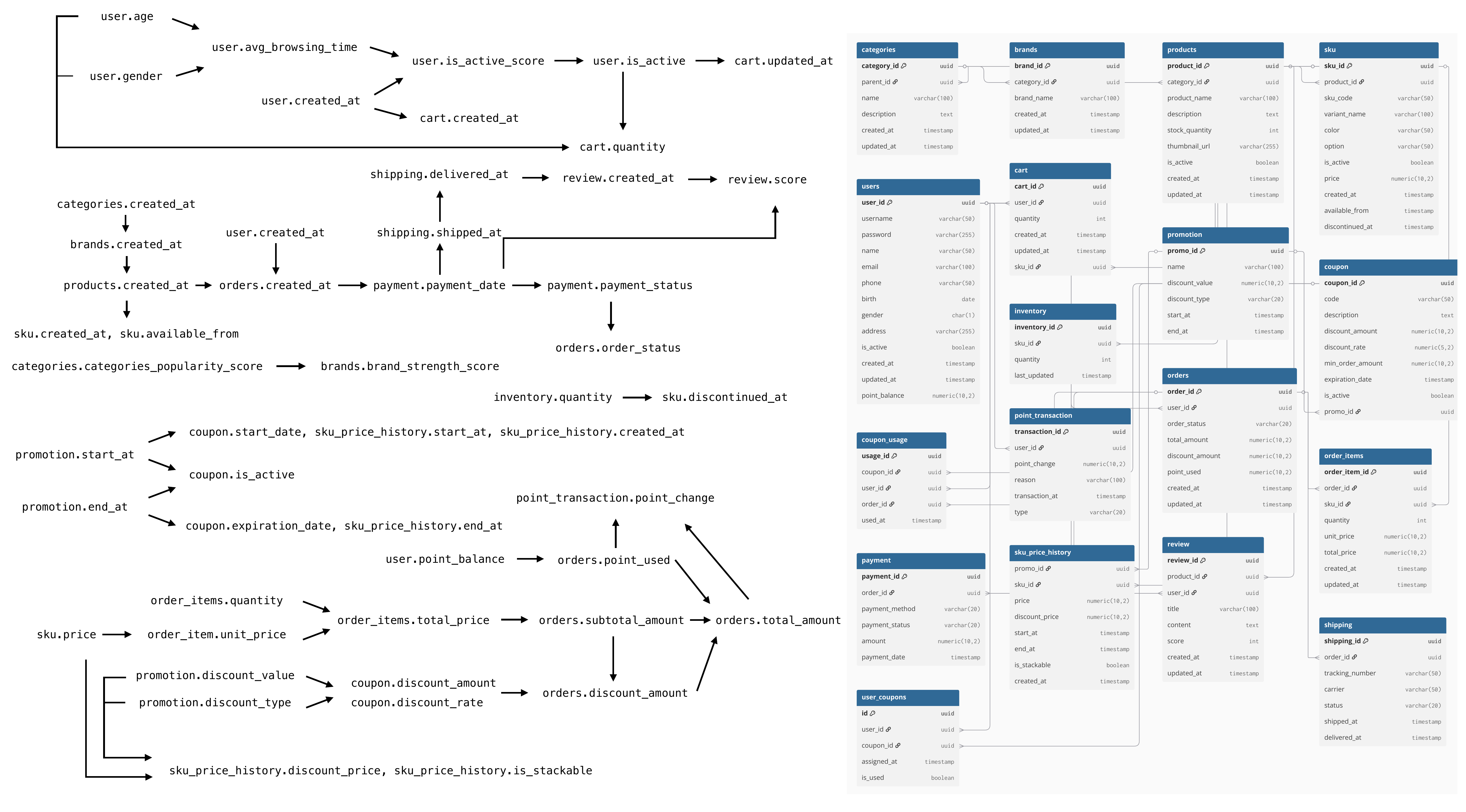}
    \caption{Causal graph and relational schema of the REEF benchmark.
    The left panel shows the ground-truth structural causal model (SCM) used to generate the data, while the right panel presents the corresponding entity–relationship diagram (ERD) of the relational database.
    Variables relevant to a single causal query are distributed across multiple tables and linked through temporal and relational dependencies.}
    \Description{The figure consists of two side-by-side panels illustrating the REEF benchmark.
    The left panel depicts a directed causal graph representing the ground-truth structural causal model.
    Nodes correspond to user attributes, behavioral signals, transactional variables, pricing and promotion factors, and post-purchase outcomes.
    Directed edges indicate causal dependencies, including temporal relationships such as account creation preceding orders, orders preceding payments and shipping, and shipping preceding reviews.
    Multiple variables influencing order totals, discounts, and user activity are shown converging into shared outcomes.
    The right panel shows the entity–relationship diagram of the corresponding relational database.
    It contains tables such as users, orders, order_items, products, brands, coupons, promotions, payments, shipping, reviews, and price history, with primary and foreign key relationships indicated by connecting lines.
    The layout highlights that variables connected in the causal graph are stored across different tables and at different granularities, requiring joins, aggregations, and temporal alignment to construct an analysis-ready dataset.
    }
    \label{fig:reef_figure}
\end{figure*}

REEF is implemented as a relational PostgreSQL database designed to reflect the structure of real-world e-commerce systems.
The schema consists of 18 tables and 130 variables spanning user profiles, product catalogs, transactional processes, and post-purchase interactions.
These tables are connected through explicit foreign-key relationships, inducing multi-hop relational paths (e.g., user $\rightarrow$ order $\rightarrow$ payment, or category $\rightarrow$ brand $\rightarrow$ product ), which are commonly encountered in enterprise analytics.
This design mirrors operational database practices and encourages evaluation under realistic analytical workflows, where relevant features must be explicitly defined through joins rather than directly observed.
Figure~\ref{fig:reef_figure} illustrates the overall relational structure and causal relationships of variables.

\subsection{Data Generation and Seeding Process}
REEF is populated using a semi-synthetic data generation pipeline that combines domain-informed structural equations with stochastic noise.
Data seeding is performed in a fixed order that respects causal and relational dependencies, starting from exogenous entities (e.g., categories, users) and progressing toward downstream transactional tables (e.g., cart, orders, payment, shipping).
Each table is generated using a combination of rule-based logic and probabilistic sampling implemented in \texttt{Faker.js}.
For example, user attributes such as age, gender, and browsing behavior are sampled from biased distributions to reflect realistic population characteristics, while latent activity scores are computed through explicit functional relationships and transformed into binary activity indicators via logistic functions.
Downstream processes, including cart quantities, order composition, discount usage, and payment outcomes, are generated conditional on upstream variables, ensuring that observed correlations arise from known causal mechanisms rather than arbitrary randomness.
Importantly, the seeding process injects controlled noise, missing values, and temporal variability (e.g., delayed payments, inactive users, discontinued products), preserving the irregularities of observational data while maintaining a known underlying data-generating process.

\subsection{Design Choices and Limitations}
REEF is designed to balance realism and causal transparency.
Rather than generating fully synthetic tabular data, we adopt a relational, event-driven design that mirrors operational databases used in practice.
This choice enables evaluation of causal methods under realistic challenges such as multi-table joins, aggregation-induced bias, and mixed variable types.
At the same time, several simplifying assumptions are made.
First, while temporal ordering is respected in data generation, REEF does not model unobserved time-varying confounders beyond those explicitly encoded in the structural equations.
Second, some treatments (e.g., categorical product attributes) require aggregation or encoding choices when used for effect estimation, which may introduce modeling ambiguity.
Finally, although missingness is intentionally introduced, it follows controlled mechanisms and does not capture all forms of non-random data absence observed in real deployments.
Despite these limitations, REEF provides a practical and reproducible benchmark that preserves the essential complexity of relational observational data while enabling rigorous evaluation against known causal ground truth.

\section{Evaluation of Specialist Agents}
\label{appendix:agent-evaluation}
\setcounter{figure}{0}
\setcounter{table}{0}
We conduct a series of experiments to evaluate the independent robustness of each specialist agents.

\subsection{Evaluation Metrics}
\label{appendix:metrics}
This section provides formal definitions of the evaluation metrics used throughout the experiments.

\paragraph{Structural Hamming Distance (SHD).}
SHD~\cite{tsamardinos2006max} measures the structural discrepancy between a predicted graph $\hat{G}$ and the ground-truth graph $G$.
It is defined as the minimum number of edge insertions, deletions, or reversals required to transform $\hat{G}$ into $G$:
\begin{equation}
\mathrm{SHD}(\hat{G}, G) =
\left| \{ (i,j) \in V^2 \mid \hat{G}_{ij} \neq G_{ij} \} \right|.
\end{equation}
Lower SHD indicates closer structural agreement.

\paragraph{Structural Intervention Distance (SID).}
SID~\cite{peters2015structural} evaluates the discrepancy between two graphs in terms of implied interventional distributions.
Formally, SID counts the number of variable pairs for which the predicted graph yields incorrect interventional effects compared to the ground-truth graph.
Lower SID indicates higher causal consistency under interventions.

\paragraph{Precision and Recall (Table Retrieval).}
For table retrieval, precision and recall are computed by comparing the set of tables recommended by the system $R_n$ with the ground-truth relevant tables $G_n$:
\begin{equation}
\mathrm{Recall} = \frac{1}{N} \sum_{n=1}^{N} \frac{|G_n \cap R_n|}{|G_n|}, \quad
\mathrm{Precision} = \frac{1}{N} \sum_{n=1}^{N} \frac{|G_n \cap R_n|}{|R_n|}.
\end{equation}

\paragraph{Execution Accuracy (EX).}
For text-to-SQL generation, Execution Accuracy measures whether the predicted SQL query produces the same execution result as the ground-truth query when run against the database:
\begin{equation}
\mathrm{EX} = \frac{1}{N} \sum_{n=1}^{N} \mathbb{1}(\hat{V}_n = V_n).
\end{equation}

\paragraph{Treatment Effect Metrics.}
For treatment effect estimation, we report:
(i) absolute error of the estimated Average Treatment Effect (ATE), (ii) bias(values closer to zero indicate better performance), (iii) confidence interval (CI) coverage, and (iv) CI width.
CI coverage is defined as the proportion of experiments in which the true effect lies within the estimated confidence interval.

\subsection{Data Exploration Agent}
\label{appendix:exploration-eval}
\begin{figure*}[t]
    \centering
    \includegraphics[width=\linewidth]{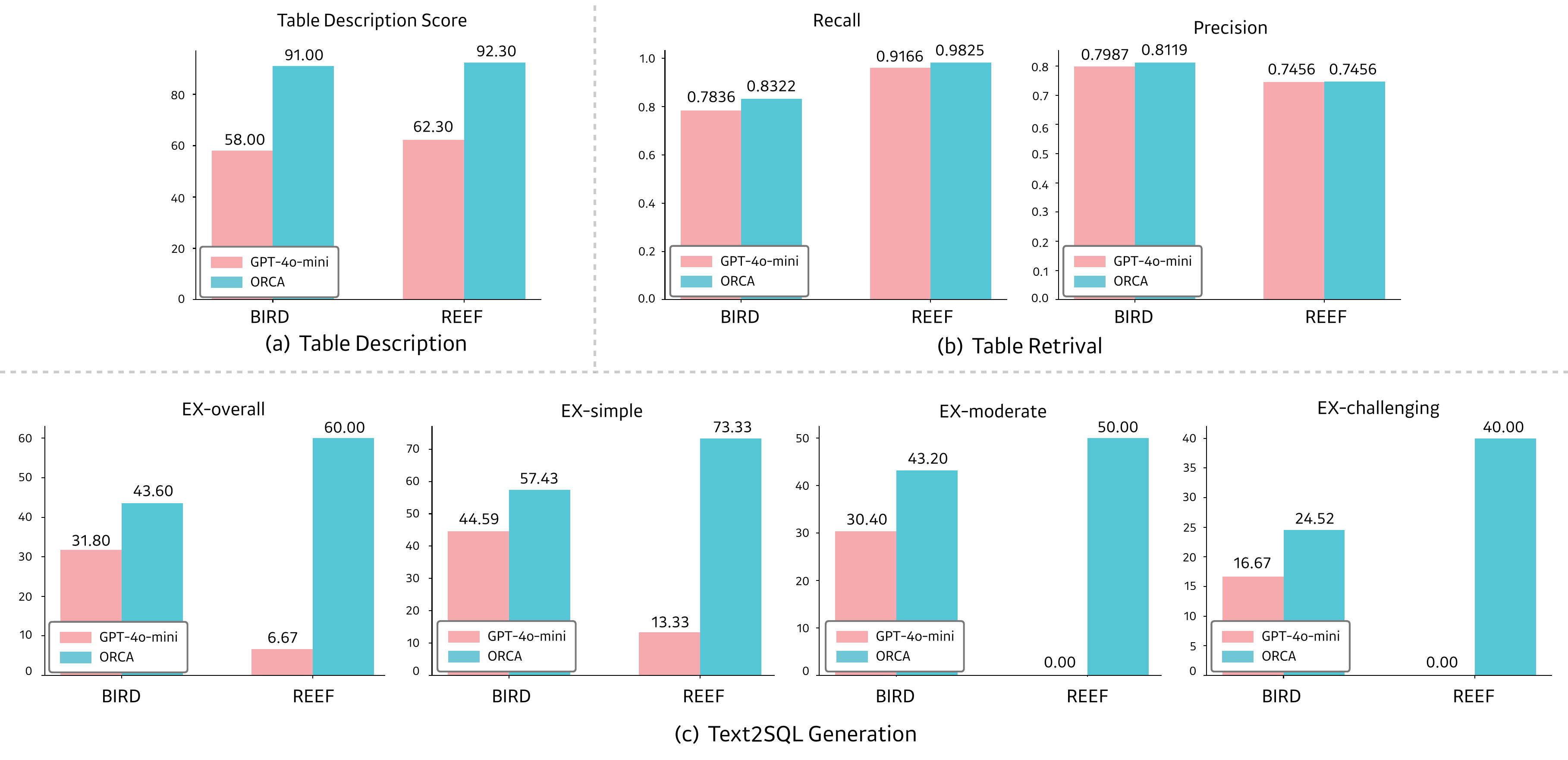}
    \caption{Evaluation results of the Data Exploration Agent across multiple subtasks.
    The figure compares ORCA and GPT-4o-mini on table description, table retrieval, and Text2SQL generation using the BIRD and REEF benchmarks.
    ORCA consistently outperforms the baseline across all subtasks, with particularly large gains on the relationally complex REEF dataset and harder Text2SQL query categories.}
    \Description{The figure contains multiple bar charts arranged in three groups, each comparing ORCA and GPT-4o-mini on different data exploration subtasks using the BIRD and REEF benchmarks.
    Panel (a) shows table description accuracy.
    For each benchmark, two bars represent the performance of GPT-4o-mini and ORCA, with ORCA achieving higher accuracy on both datasets.
    Panel (b) shows table retrieval performance, reported using recall and precision.
    Separate bar charts compare the two models on BIRD and REEF, with ORCA consistently achieving higher recall and precision.
    Panel (c) shows Text2SQL execution accuracy, broken down by query difficulty levels.
    Bar charts compare the two models on BIRD and REEF across increasing levels of query complexity.
    Performance gaps widen on harder queries, especially on the REEF benchmark, where ORCA maintains high execution accuracy while GPT-4o-mini degrades substantially.
    Across all panels, ORCA outperforms the baseline model, with the largest improvements observed on tasks requiring multi-table reasoning and complex relational understanding.
    }
    \label{fig:DataExploration_result}
\end{figure*}

\paragraph{Dataset}
We use REEF and BIRD~\cite{Li2023CanLA} data to evaluate \textbf{Data Exploration Agent}.
BIRD is a text-to-SQL benchmark grounded in real-world domains, designed to test models under challenging conditions such as external knowledge reasoning, SQL efficiency over large databases, and complex query structures including implicit joins and aggregations. 
For reproducibility and efficient error analysis, we adopt the official BIRD mini-dev subset, which includes 500 natural language queries across 12 diverse databases\footnote{\url{https://github.com/bird-bench/mini_dev}}. 

\subsubsection{Table Description Evaluation.}
We evaluate the ability of \textit{Table Explorer} to generate informative and concise natural language descriptions of database tables.
For each table, both the baseline and ORCA are tasked with producing a description that conveys the core semantics and structure of the table. 
The quality of these descriptions is then assessed using a standardized rubric.

The baseline model is prompted to take a table as input and to create a description.
Due to the large size of the tables, only the first 100 rows are used as input for baseline model. 
The evaluation is conducted using a 5-point scale rescaled to 0--100 for reporting, which measures how informative and meaningful each description is. 
We note that GPT-4o-mini is used as an automatic evaluator, which may introduce model-based evaluation bias.
Accordingly, the reported scores are intended for relative comparison between methods under identical conditions, rather than as absolute measures of quality.
To ensure fairness, GPT-4o-mini is prompted with clear and consistent grading instructions.
Detailed grading criteria are provided in the Appendix~\ref{appendix-sec:prompts}.

As shown in Figure~\ref{fig:DataExploration_result}(a), ORCA significantly outperforms the baseline in both datasets, achieving average scores of 91.0 and 92.3 compared to 58.0 and 62.3, respectively. 
This demonstrates ORCA’s strong ability to interpret and summarize structured data effectively.

\subsubsection{Table Retrieval Evaluation.} 
We also evaluate the ability of the \textit{Table Recommender} to recommend relevant tables given a natural language question. 
The goal of this task is to identify which tables from a database are necessary to answer a user query.
The baseline model (GPT-4o-mini) is prompted with a list of all table names in the database, followed by the user question. 
It is instructed to select which tables would be helpful for answering the question. 
We evaluate table retrieval using Recall and Precision, measuring coverage of ground-truth tables and correctness of recommended tables, respectively.
As shown in Figure~\ref{fig:DataExploration_result} (b), ORCA consistently outperforms baseline on both datasets, achieving higher Recall and Precision. 
This indicates that ORCA is more effective at identifying relevant tables while minimizing the inclusion of unnecessary ones.

\subsubsection{Text2SQL Generator Evaluation.}
We evaluate the ability of \textit{Text2SQL Generator} to generate a valid SQL query for the user's request.
Each SQL prediction is executed against the corresponding database, and the output is compared with that of the ground-truth SQL to compute execution accuracy. 
Execution Accuracy (EX) measures the percentage of evaluation examples where the predicted SQL and the ground-truth SQL produce the same execution result. 
As shown in Figure~\ref{fig:DataExploration_result} (c), ORCA consistently outperforms the baseline across all difficulty levels on both datasets. 
Notably, while the REEF dataset reflects a more realistic and structurally complex database environment, ORCA shows a strong execution accuracy of 60.00\%, whereas GPT-4o mini’s performance significantly drops to 6.67\%.

\subsubsection{Comparison with Frontier LLMs.}
To verify that ORCA's improvements are not simply replicated by adopting a stronger backbone LLM, we evaluate GPT-5 and Gemini-2.5-Flash on all three Data Exploration tasks using the REEF dataset. 
For fair comparison, the same model is used as both the standalone baseline and ORCA's internal LLM.
As shown in Table~\ref{tab:de_results_models}, frontier baselines improve Table Explorer scores over GPT-4o-mini (see Fig.~\ref{fig:DataExploration_result}), yet ORCA achieves a perfect score with both backbones.
For Table Recommender, ORCA shows stronger recall, suggesting it is more reliable at surfacing all relevant tables for a given query.
On Text2SQL, ORCA consistently outperforms each corresponding baseline, though absolute gains remain modest given the inherent difficulty of SQL generation on complex schemas.
Together with the causal discovery results in Table~\ref{tab:cd_results_all}, these findings indicate that ORCA's agentic workflow adds value beyond what stronger LLMs alone provide.

\begin{table*}[t]
\centering
\small
\begin{tabular}{llcccccc}
\toprule
& & \textbf{Table Explorer} & \multicolumn{3}{c}{\textbf{Table Recommender}} & \textbf{Text2SQL} \\
\cmidrule(lr){3-3} \cmidrule(lr){4-6} \cmidrule(lr){7-7}
\textbf{Backbone LLM} & \textbf{Method}
& Quality Score $\uparrow$
& Precision $\uparrow$ & Recall $\uparrow$ & F1 $\uparrow$
& EX $\uparrow$ \\
\midrule

\multirow{2}{*}{GPT-5}
& Baseline & 91.2 & \textbf{0.766} & 0.868 & 0.796 & 0.467 \\
& ORCA     & \textbf{100.00} & 0.599 & \textbf{1.000} & 0.723 & \textbf{0.500} \\
\midrule

\multirow{2}{*}{Gemini-2.5-Flash}
& Baseline & 82.2 & \textbf{0.730} & 0.807 & 0.748 & 0.433 \\
& ORCA     & \textbf{100.00} & 0.754 & \textbf{0.842} & \textbf{0.770} & \textbf{0.467} \\

\bottomrule
\end{tabular}
\caption{Data Exploration module performance on the REEF dataset with frontier LLMs. For fair comparison, the same backbone LLM is used both as the standalone baseline and as ORCA's internal model.}
\label{tab:de_results_models}
\end{table*}

\subsection{Causal Discovery Agent}
\label{appendix:discovery-eval}
\paragraph{Dataset} We evaluate causal discovery performance using synthetic datasets generated from random directed acyclic graphs (DAGs) under controlled data-generating processes.
As summarized in Table~\ref{tab:cd_synthetic_settings}, we consider five scenarios that vary graph topology (Erd\H{o}s--R\'enyi vs.\ scale-free), SCM functional form (linear vs.\ nonlinear), and noise distribution (Gaussian vs.\ non-Gaussian).
Given a ground-truth DAG, data are generated in topological order using additive SCMs, where each variable is computed as a normalized sum of its parent variables followed by either a linear transformation or a nonlinear activation, and additive noise.
Noise terms are sampled independently from either a Gaussian or Laplace distribution, depending on the scenario.
Across all settings, the number of variables is varied as $d \in \{3,5,10\}$, and sample sizes are chosen such that $N > D$.
For each scenario and dimensionality, we generate 10 independent datasets using different random seeds.

\begin{table*}[t]
\centering
\small
\resizebox{\linewidth}{!}{
\begin{tabular}{lcccc}
\toprule
\textbf{Scenario} & \textbf{Graph} & \textbf{SCM} & \textbf{Noise} & \textbf{Scale} \\
\midrule
Baseline (ER--Linear, Gaussian) 
& ER & Linear & Gaussian & $N > D$ \\

Graph Complexity (Scale-Free--Linear) 
& SF & Linear & Gaussian & $N > D$ \\

Noise Misspecification (ER--Linear, Non-Gaussian) 
& ER & Linear & Non-Gaussian (Laplace) & $N > D$ \\

Structural Misspecification (ER--Nonlinear) 
& ER & Non-linear (e.g., MLP) & Gaussian & $N > D$ \\

Fully Challenging Setting (Scale-Free--Nonlinear) 
& SF & Non-linear & Non-Gaussian & $N > D$ \\
\bottomrule
\end{tabular}
}
\caption{Synthetic causal discovery scenarios.}
\label{tab:cd_synthetic_settings}
\end{table*}

\paragraph{Setting} Given observational data sampled from a known ground-truth DAG, each method estimates a directed causal graph.
We compare ORCA against two baselines: AutoCD, a classical automated causal discovery framework, and GPT-4o-mini, an LLM-based baseline that directly infers causal relations from data summaries; and two frontier LLMs, GPT-5 and Gemini-2.5-Flash, evaluated under the same prompting protocol as GPT-4o-mini.
All methods are evaluated under identical data-generating conditions, and performance is averaged across repeated runs.

\paragraph{Results} Table~\ref{tab:cd_results_all} reports causal discovery performance across synthetic scenarios.
In most settings, ORCA achieves lower structural errors than all baselines.
Notably, while frontier models occasionally improve over GPT-4o-mini on individual scenarios (e.g., Gemini-2.5-Flash on ER--Linear at $d{=}10$), ORCA achieves the most consistent gains across graph topologies and functional forms.
In scale-free graphs, which better reflect real-world causal sparsity patterns, ORCA achieves large and consistent improvements over all baselines.
Also, ORCA shows robustness in performance even with noise perturbations or complex dependency structures.
These results indicate that ORCA successfully guides causal discovery across varying graph topologies and functional forms, especially in settings where structural assumptions are difficult to specify a priori.
For higher-dimensional setting with $d=10$ variables, we restrict this analysis to three representative scenarios due to the rapidly increasing search space: ER--Linear, Scale-Free--Linear, and ER--Nonlinear.
In Scale-Free--Linear and ER--Nonlinear scenarios, ORCA achieves lower SHD and SID than both baselines, indicating improved resilience to increased graph complexity and nonlinear mechanisms even in higher dimensions.
AutoCD outputs do not directly support SID computation due to the absence of a fully specified directed graph; therefore, SID is omitted for this method.

\begin{table*}[t]
\centering
\small
\begin{tabular}{llcccccc}
\toprule
\textbf{Scenario} & \textbf{Method}
& \multicolumn{2}{c}{$d=3$}
& \multicolumn{2}{c}{$d=5$}
& \multicolumn{2}{c}{$d=10$} \\
\cmidrule(lr){3-4} \cmidrule(lr){5-6} \cmidrule(lr){7-8}
& 
& \textbf{SHD $\downarrow$} & \textbf{SID $\downarrow$}
& \textbf{SHD $\downarrow$} & \textbf{SID $\downarrow$}
& \textbf{SHD $\downarrow$} & \textbf{SID $\downarrow$} \\
\midrule

\multirow{5}{*}{ER--Linear}
& AutoCD              & 3.0 & --   & 10.0 & --   & 41.4 & --   \\
& GPT-4o-mini          & 3.2 & 4.4  & 11.3 & 15.8 & \textbf{40.8} & \textbf{81.1} \\
& GPT-5.2                & 3.4 & 4.6  & 10.5 & 16.2 & 40.4 & 79.1 \\
& Gemini-2.5-Flash     & 3.5 & 4.4  & 9.4  & 14.5 & 39.8 & 76.8 \\
& ORCA                 & \textbf{1.9} & \textbf{2.1} & \textbf{10.8} & \textbf{15.1} & 44.2 & 83.3 \\
\midrule

\multirow{5}{*}{Scale-Free--Linear}
& AutoCD              & 2.0 & --   & 4.2  & --   & 25.4 & --   \\
& GPT-4o-mini          & 3.3 & 4.2  & 8.0  & 11.5 & 24.8 & 67.2 \\
& GPT-5.2                & 2.4 & 3.6  & 6.0  & 10.9 & 27.6 & 68.8 \\
& Gemini-2.5-Flash     & 2.4 & 3.6  & 7.7  & 14.4 & 29.5 & 66.8 \\
& ORCA                 & \textbf{0.2} & \textbf{0.3} & \textbf{1.1} & \textbf{2.0} & \textbf{21.4} & \textbf{58.6} \\
\midrule

\multirow{5}{*}{ER--Linear--NonGaussian}
& AutoCD              & 3.0 & --   & 10.0 & --   & --   & --   \\
& GPT-4o-mini          & \textbf{2.0} & 2.6  & 10.0 & 16.8 & --   & --   \\
& GPT-5.2                & 2.4 & 2.6  & 9.3  & 15.8 & --   & --   \\
& Gemini-2.5-Flash     & 2.5 & 3.5  & 8.6  & 14.7 & --   & --   \\
& ORCA                 & \textbf{2.0} & \textbf{2.2} & \textbf{7.9} & \textbf{12.0} & --   & --   \\
\midrule

\multirow{5}{*}{ER--Nonlinear}
& AutoCD              & 3.0 & --   & 10.0 & --   & 39.3 & --   \\
& GPT-4o-mini          & 2.5 & 2.9  & \textbf{10.1} & \textbf{15.3} & 41.2 & 84.6 \\
& GPT-5.2                & 2.6 & 3.4  & 10.2 & 15.2 & 40.6 & 78.1 \\
& Gemini-2.5-Flash     & 2.3 & 3.2  & 10.0 & 15.5 & 42.4 & 80.6 \\
& ORCA                 & \textbf{1.4} & \textbf{1.6} & 10.4 & 16.3 & \textbf{38.9} & \textbf{76.5} \\
\midrule

\multirow{5}{*}{Scale-Free--Nonlinear}
& AutoCD              & 2.2 & --   & 4.6  & --   & --   & --   \\
& GPT-4o-mini          & 2.0 & 2.4  & 5.3  & 6.3  & --   & --   \\
& GPT-5                & 1.6 & 2.4  & 4.5  & 6.4  & --   & --   \\
& Gemini-2.5-Flash     & 1.6 & 2.4  & 5.9  & 6.7  & --   & --   \\
& ORCA                 & \textbf{0.6} & \textbf{0.9} & \textbf{1.7} & \textbf{3.7} & --   & --   \\

\bottomrule
\end{tabular}
\caption{Causal discovery performance on synthetic datasets at $d=3$, $d=5$, and $d=10$ (mean over 10 runs).}
\label{tab:cd_results_all}
\end{table*}

\subsection{Causal Inference Agent}
\label{appendix:inference-eval}
\paragraph{Dataset} To evaluate treatment effect estimation performance of the Causal Inference Agent, we use both synthetic data tables and REEF with complex, multi-table data structures.
Synthetic datasets are designed to evaluate robustness to confounding, model misspecification, and heterogeneous treatment effects.
As summarized in Table~\ref{tab:synthetic_settings}, we consider four scenarios that vary the functional form of treatment assignment, outcome generation, and the presence of heterogeneous treatment effects.
Confounders are sampled from a standard normal distribution, and treatment is assigned via a logistic propensity model that is either linear or nonlinear in the confounders.
Potential outcomes are generated using additive outcome models with either constant(ATE) or confounder-dependent (CATE) treatment effects, and the observed outcome is constructed according to the assigned treatment.
Nonlinear outcome scenarios introduce misspecification through higher-order confounder terms.
All treatment effect datasets use a fixed sample size ($N=2000$) and number of confounders ($D=5$), and each scenario is generated with multiple random seeds.

\begin{table*}[t]
\centering
\resizebox{\linewidth}{!}{
\begin{tabular}{cllll}
\toprule
\textbf{Scenario} & \textbf{Treatment Assignment} & \textbf{Outcome Function} & \textbf{Effect Type} & \textbf{Purpose} \\
\midrule
Baseline 
& Linear $P(T \mid W)$ 
& Linear $E[Y \mid T, W]$ 
& ATE 
& Baseline sanity check \\

Non-linear Treatment Assignment 
& Non-linear (e.g., $W^2$) 
& Linear $E[Y \mid T, W]$ 
& ATE 
& Robustness to non-linear assignment \\

Non-linear Outcome Model
& Linear $P(T \mid W)$ 
& Non-linear (e.g., $W^2$) 
& ATE 
& Outcome model misspecification \\

Heterogeneous Treatment Effects
& Linear $P(T \mid W)$ 
& Interaction $(T \times W)$ 
& CATE 
& Heterogeneous treatment effects \\

\bottomrule
\end{tabular}
}
\caption{Synthetic data-generating settings for treatment effect estimation experiments.}
\label{tab:synthetic_settings}
\end{table*}

\paragraph{Setting} To distinguish estimation error caused by incorrect causal structure from error arising from statistical estimation, we explicitly use two experimental settings.
In the Oracle Graph setting, estimators are provided with the ground-truth causal graph.
This setting isolates the pure estimation capability of each method, independent of any error in causal discovery.
In the Agentic Graph setting, effect estimation is performed using causal graphs inferred by ORCA or baseline discovery methods, allowing to assess how discovery-stage errors propagate to downstream effect estimation.
For Oracle Graph settings, we compare ORCA with standard baselines including DoubleML~\cite{bach2022doubleml}, Causal Forest~\cite{wager2018estimation}, and the T-learner~\cite{kunzel2019metalearners}.
The Agentic Graph setting applies only to methods that output explicit causal graphs; accordingly, this setting evaluates effect estimation using graphs inferred by ORCA.

\paragraph{Results} Table~\ref{tab:effect_estimation_synthetic} summarizes treatment effect estimation results across four synthetic scenarios under the oracle graph setting.
Across all scenarios, ORCA consistently achieves the lowest ATE absolute error, ranging from 0.0030 in the linear baseline to 0.0077 in the heterogeneous treatment effect setting, outperforming all baseline estimators.
In the non-linear outcome setting, ORCA remains stable while competing methods exhibit degraded coverage.
ORCA also maintains high CI coverage (80–100\%) while producing relatively narrower confidence intervals.

Table~\ref{tab:effect_estimation_reef} reports results on the REEF benchmark.
Under the oracle graph setting, ORCA attains near-zero ATE error ($3.2\times10^{-5}$) with 100\% CI coverage, significantly outperforming strong baselines such as DoubleML and Causal Forest.
When causal graphs are inferred rather than provided, estimation accuracy degrades as expected due to error propagation(e.g. missing confounders) from the discovery stage.
Nevertheless, under the agentic graph setting, ORCA preserves high CI coverage and success rate on REEF, despite increased ATE error, demonstrating robustness to graph uncertainty.

\begin{table*}[t]
\centering
\small
\begin{tabular}{llcccc}
\toprule
\textbf{Synthetic scenario} & \textbf{Method} & \textbf{ATE Abs Error $\downarrow$} & \textbf{ATE Bias $\downarrow$} & \textbf{CI Coverage (\%) $\uparrow$} & \textbf{CI Width $\downarrow$} \\
\midrule

\multirow{4}{*}{\textbf{Linear Baseline}}
& Causal Forest & 0.0283 & -0.0283 & \textbf{100.0} & 0.1620 \\
& DoubleML      & 0.0113 & -0.0067 & \textbf{100.0} & 0.0585 \\
& T-learner     & 0.1057 &  0.1057 & ---   & ---    \\
& ORCA (ours)   & \textbf{0.0030} & \textbf{-0.0018} & \textbf{100.0} & \textbf{0.0197} \\
\midrule

\multirow{4}{*}{\textbf{Non-linear Treatment Assignment}}
& Causal Forest & 0.0289 & -0.0285 & \textbf{100.0} & 0.1740 \\
& DoubleML      & 0.0321 & -0.0321 & 40.0  & 0.0581 \\
& T-learner     & 0.0643 &  0.0643 & ---   & ---    \\
& ORCA (ours)   & \textbf{0.0053} & \textbf{0.0025} & 80.0 & \textbf{0.0192} \\
\midrule

\multirow{4}{*}{\textbf{Non-linear Outcome Model}}
& Causal Forest & 0.0392 & -0.0392 & \textbf{100.0} & 0.0414 \\
& DoubleML      & 0.0267 & -0.0252 & 40.0  & \textbf{0.0306} \\
& T-learner     & 0.0949 &  0.0949 & ---   & ---    \\
& ORCA (ours)   & \textbf{0.0250} & \textbf{-0.0036} & \textbf{100.0} & 0.0349 \\
\midrule

\multirow{4}{*}{\textbf{Heterogeneous Treatment Effects (CATE)}}
& Causal Forest & 0.0125 & -0.0044 & \textbf{100.0} & 0.2123 \\
& DoubleML      & 0.0157 & \textbf{-0.0013} & \textbf{100.0} & 0.0746 \\
& T-learner     & 0.1326 &  0.1326 & ---   & ---    \\
& ORCA (ours)   & \textbf{0.0077} & -0.0032 & \textbf{100.0} & \textbf{0.0529} \\
\bottomrule
\end{tabular}
\caption{Treatment effect estimation performance on synthetic scenarios (oracle graph setting).}
\label{tab:effect_estimation_synthetic}
\end{table*}

\begin{table*}[t]
\centering
\begin{tabular}{llccccc}
\toprule
\textbf{Graph Setting} & \textbf{Method} 
& \textbf{ATE Abs Error $\downarrow$} 
& \textbf{ATE Bias $\downarrow$} 
& \textbf{CI Coverage (\%) $\uparrow$} 
& \textbf{CI Width $\downarrow$} 
& \textbf{Success rate (\%) $\uparrow$} \\
\midrule

\multirow{4}{*}{\textbf{Oracle Graph}}
& Causal Forest & 0.0595 & -0.0571 & 88 & 1.1089 & \textbf{100} \\
& DoubleML      & 0.0527 & 0.0020 & 84 & 1.0294 & \textbf{100} \\
& T-learner     & 0.0270 & -0.0071 & --- & --- & 24 \\
& ORCA (ours)   & \textbf{0.000032} & \textbf{0.000032} & \textbf{100} & 1.6338 & \textbf{100} \\
\midrule

\multirow{1}{*}{\textbf{Agentic Graph}}
& ORCA (ours)   & 0.0414 & -0.0407 & 88 & 1.0531 & 100 \\
\bottomrule
\end{tabular}
\caption{Treatment effect estimation performance on the REEF benchmark.}
\label{tab:effect_estimation_reef}
\end{table*}

\section{User Study Details}
\label{appendix:user_study}
\setcounter{figure}{0}
\setcounter{table}{0}

\subsection{Detailed Study Protocol}
\label{appendix:user_study:protocol}
This appendix provides the full study protocol used to evaluate ORCA against a LLM-assisted coding baseline for end-to-end causal analysis on a relational database benchmark (REEF).
We focus on a single relational benchmark (REEF) to ensure a controlled causal setting with known ground-truth effects, allowing performance differences to be attributed to the analysis workflow rather than dataset-specific confounds.
In the absence of a formal institutional review process, we followed internationally accepted ethical principles, including informed consent, voluntary participation, and data anonymization.
All participants were informed about the study purpose and data usage and provided consent prior to participation.
Participants were informed that interaction logs would be collected for research purposes, and consented to this data collection as part of the study agreement.
The study was conducted remotely and followed a within-subjects design in which each participant completed both conditions.
To mitigate order effects, participants were randomly assigned to one of two counterbalanced groups:
(i) Baseline $\rightarrow$ ORCA, or (ii) ORCA $\rightarrow$ Baseline.

\subsubsection{Timeline and Phases}
The session was designed to take approximately 120 minutes in total, consisting of three phases:

\begin{itemize}
    \item \textbf{Phase 1: Pre-setting (20 min)}
    Participants received study instructions and a short tutorial covering database setup and how to operate each system (Baseline and ORCA), including how to save and submit logs.
    Participants then completed a pre-study survey (approximately 5 minutes) assessing demographics, programming experience (Python/SQL), debugging familiarity, prior experience with LLM-assisted coding tools, and causal inference knowledge.
    
    \item \textbf{Phase 2: Main Sessions (up to 60 min $\times$ 2)}
    Participants completed two analysis tasks under two different conditions (Baseline vs.\ ORCA), following the assigned counterbalanced order.
    Each task had a strict time limit of \textbf{60 minutes}.
    If the time limit was exceeded, participants stopped the current task and proceeded to the next session.
    
    \item \textbf{Phase 3: Post-experiment Survey (10 min)}
    After completing both conditions, participants answered an overall post-experiment questionnaire including comparative items (ORCA vs.\ Baseline) and optional open-ended feedback.
\end{itemize}

\subsubsection{Task Pipeline}
Both conditions followed the same three-step pipeline and required participants to submit step-wise artifacts:
\begin{enumerate}
    \item \textbf{Step 1: Data exploration and dataset retrieval (Data Exploration)}
    Participants explored the database schema and identified the tables, columns, join paths, and variable definitions required to answer the causal question.
    \textbf{Artifacts:} (i) final SQL query (\texttt{sql}), and (ii) final analysis dataset (\texttt{data.parquet}).
    
    \item \textbf{Step 2: Causal structure construction (Causal Discovery)}
    Participants specified or refined a causal graph including the treatment--outcome relation and relevant variables (e.g., confounders, mediators, colliders).
    \textbf{Artifacts:} final causal graph (\texttt{graph.json}).
    
    \item \textbf{Step 3: Effect estimation and interpretation (Causal Inference)}
    Participants estimated the average treatment effect (ATE) and provided a brief interpretation, including uncertainty checks when applicable.
    \textbf{Artifacts:} estimation output (\texttt{ate.json}).
\end{enumerate}

\subsubsection{Operational Rules and Data Collection}
Throughout the study, participants were instructed to externalize their reasoning (think-aloud) by briefly stating what they were attempting and why (or recording short notes).
For ORCA, system logs were collected automatically and submitted as a compressed archive.
For the Baseline condition, the interaction history was recorded via screen capture and/or saved prompts and code, following the same artifact submission structure whenever possible.
Participants additionally reported whether they completed each task within the time limit and submitted the final log archive for each run.

\subsection{Baseline Implementation}
The baseline is implemented as a single, conversation-based LLM assistant with step-gated progression.
Within each step, participants interact through free-form dialogue and tool calls for schema inspection, SQL execution, Python-based analysis, and artifact saving.
Step transitions are explicitly controlled by a fixed user command (\texttt{/next}), which triggers a finalization routine that requires saving step-specific artifacts (Step1: SQL query and dataset; Step2: causal graph or adjacency representation; Step~3: ATE estimate), as specified in the prompt used for the baseline shown in Figure~\ref{fig:baseline_prompt}.

To reflect a basic architecture capable of maintaining continuity across interactions, the baseline is designed to preserve state across steps.
Rather than relying on long conversation histories, it maintains a persistent session state (\texttt{state.json}) that records artifact paths and key variables (e.g., treatment, outcome, confounders), and re-injects a compact summary of this state as system context at the start of each step.
Step completion is programmatically validated by checking the presence of required artifact types before advance.
Overall, this design yields a reproducible, tool-backed workflow with minimal but explicit state discipline, comparable to common LLM-based coding assistants augmented with execution tools.

\subsection{Participants}
\begin{figure*}[t]
    \centering
    \includegraphics[
        width=\linewidth,
        height=0.85\textheight,
        keepaspectratio
    ]{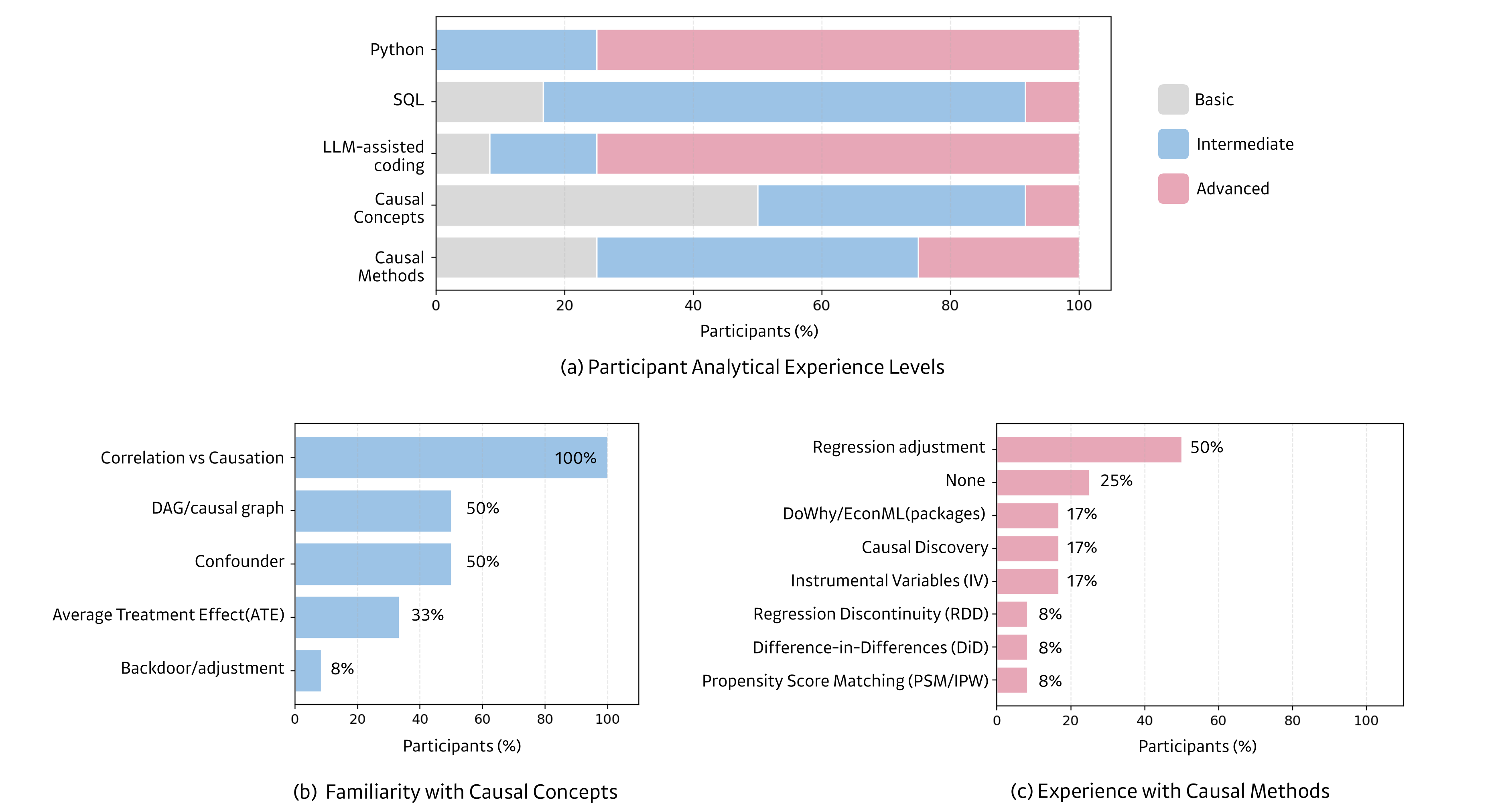}
    \caption{Overview of participants’ analytical background and causal experience ($N=12$).
    (a) Distribution of participants’ self-reported analytical experience across core skills, including Python, SQL, LLM-assisted coding, and causal analysis, categorized as basic, intermediate, or advanced.
    (b) Proportion of participants reporting familiarity with causal inference concepts.
    (c) Proportion of participants reporting prior hands-on experience with specific causal inference methods.
    Overall, the distributions indicate that participants were generally experienced in data analysis and familiar with core causal concepts, whereas fewer reported direct experience with advanced causal inference methods.}
    \Description{Overview of participants’ analytical background and causal experience ($N=12$).
    (a) Distribution of participants’ self-reported analytical experience across core skills, including Python, SQL, LLM-assisted coding, and causal analysis, categorized as basic, intermediate, or advanced.
    (b) Proportion of participants reporting familiarity with causal inference concepts.
    (c) Proportion of participants reporting prior hands-on experience with specific causal inference methods.
    Overall, the distributions indicate that participants were generally experienced in data analysis and familiar with core causal concepts, whereas fewer reported direct experience with advanced causal inference methods.}
    \label{fig:participant_profile}
\end{figure*}

The study involved $N=12$ participants with prior experience in data analysis, including undergraduate students (7), master’s students (4), and early-career practitioners (1), drawn from statistics, data science, computer science, and business-related fields.
Participants were generally data-literate and reported substantial experience with Python-based analysis, SQL querying, and LLM-assisted coding tools.
Figure~\ref{fig:participant_profile} provides a detailed characterization of participants’ analytical experience and causal background.
As shown in Figure~\ref{fig:participant_profile}(a), most participants reported intermediate to advanced experience in core analytical skills, reflecting familiarity with common data processing and analysis workflows.
Participants also regularly used LLMs to support tasks such as code generation, SQL writing, and exploratory data analysis.
While familiarity with high-level causal concepts was widespread (Figure~\ref{fig:participant_profile}(b)), hands-on experience with formal causal inference methods was less common (Figure~\ref{fig:participant_profile}(c)).
In particular, methods requiring specialized identification strategies such as instrumental variables, difference-in-differences, or regression discontinuity designs were reported by only a small subset of participants.

Overall, this participant profile reflects realistic data analysts who routinely work with complex relational data and modern analysis tools, but who often lack extensive training or practical experience in formal causal inference methodologies.

\subsection{Quantitative Results}
\label{appendix:user_study:quant}

\subsubsection{Metric} 
We report quantitative outcomes for both efficiency and correctness.
All metrics were computed per participant, per condition, and per task, and summarized using mean$\pm$standard deviation (or median and interquartile range when appropriate).

\begin{itemize}
    \item \textbf{Task Completion}: A run is considered complete if an ATE estimate is successfully produced, indicating that the full analysis pipeline was executed end-to-end, regardless of estimation accuracy.
    \item \textbf{Discovery Success}: A run is marked successful if the inferred causal graph contains both the treatment and outcome nodes.
    \item \textbf{Data Selection Success}: A run is marked successful if the final analysis dataset includes both the treatment and outcome variables.
    \item \textbf{CI Coverage}: Counted when the estimated ATE falls within the confidence interval of the ground-truth effect.
    \item \textbf{ATE Error}: The absolute difference between the estimated ATE and the ground-truth ATE.
    \item \textbf{Normalized Hamming Distance (NHD)}: The Hamming distance between the inferred and ground-truth causal graphs, normalized by the maximum possible number of directed edges. 
    Lower values indicate greater structural similarity.
    \item \textbf{User Interaction Time}: The total duration of active user engagement excluding tool execution time. 
    Extreme values are clipped at 0.9 to mitigate the impact of device-level anomalies such as local input/output delays.
    \item \textbf{User Interaction Turns}: The number of additional, non-mandatory user interactions beyond the minimum required workflow, capturing trial-and-error behavior.
\end{itemize}

\subsubsection{Results}

Table~\ref{tab:user_study_subfigure} summarizes user study outcomes across baseline and ORCA conditions.
Across all reported metrics, ORCA consistently outperforms the baseline.
Specifically, ORCA achieves higher task completion and causal discovery success rates (0.83 vs. 0.75 and 0.83 vs. 0.67, respectively), attains correct dataset for analysis (1.0), and improves confidence interval (CI) coverage (0.50 vs. 0.083), indicating more reliable causal effect estimation outcomes.

Table~\ref{tab:user_interaction} reports user interaction time and interaction turns across analysis stages.
During the data exploration stage, which dominates overall analysis time in relational causal workflows, ORCA significantly reduces user interaction time (80.60s vs. 439.34s on average).
Despite this reduction, ORCA maintains comparable or fewer non-mandatory interaction turns across all stages, suggesting that efficiency gains arise from reduced trial-and-error rather than increased user effort.

\subsubsection{Validation of Results}

\begin{table}[h] 
\centering
\begin{tabular}{lcc}
\toprule
\textbf{Metric} & \textbf{$\beta_{\text{cond}}$} & \textbf{\textit{p}} \\
\midrule
Data Exploration Step (s)  & $-675.49$ & $.029^*$ \\
Causal Discovery Step (s)  & $+108.34$ & $.070$   \\
Effect Estimation Step (s) & $+193.28$ & $.128$   \\
Graph Structural Error (NHD)   & $-0.0079$ & $< .001^*$ \\
ATE Absolute Error (w/ outliers) & $-16.76$ & $.321$ \\
ATE Absolute Error (w/o outliers)& $-0.0598$ & $.020^*$ \\
\bottomrule
\end{tabular}
\caption{Validation results for each result.
 $\beta_{\text{cond}}$ represents the estimated difference between ORCA and baseline, controlling for task difficulty and individual differences.}
\label{tab:validation_result}
\end{table}

To rigorously validate our findings under a small-sample (N=12) study design, we fitted linear mixed models (LMMs) for each metric with condition (ORCA vs. baseline) and task as fixed effects and participant as a random intercept, controlling for task difficulty and individual differences. 
Table ~\ref{tab:validation_result} reports the estimated condition effects; we use $p = .05$ as the significance threshold.

ORCA significantly reduced data exploration time by an estimated 675 seconds ($p = .029$). 
The causal discovery and effect estimation steps showed non-significant trends toward modestly increased time ($p = .070$, $p = .128$).
This aligns with ORCA's design intent: by compressing repetitive data exploration, users reallocated effort toward focused causal reasoning.
For graph structural error, NHD was lower under ORCA ($p < .001$), confirming improved causal structure learning. 
For ATE absolute error, the full-data model yielded a non-significant result($\beta = -16.76$, $p = .321$) due to three participants who misconfigured the analysis dataset and produced errors an order of magnitude larger than the typical range.
After excluding these task-configuration errors as outliers (IQR × 1.5), residual variance dropped substantially and the effect became significant under OLS ($\beta = -0.0598$, $p = .020$). 
We report both results for transparency.

\subsection{Survey Questions and Responses}
\label{appendix:user_study:survey}

We collected both structured ratings and open-ended responses.
The survey included a pre-study questionnaire (background and prior experience), post-task ratings (per condition), and a post-experiment questionnaire comparing ORCA against the Baseline.

\subsubsection{Question List}
\label{appendix:user_study:survey:questions}

\paragraph{Pre-study questionnaire (background)}
Items included: demographics (education level, age range, gender, and field of study), primary data analysis context, Python/SQL experience, debugging familiarity, experience with LLM-assisted coding (e.g., Code Interpreter), and prior causal inference exposure (learning experience, familiarity with concepts, and methods previously used).

\paragraph{Post-task ratings (1--7 Likert)}
Participants rated the system after each task along the following constructs:

\begin{itemize}
    \item \textbf{Usability / Ease of Use:}
    (i) ``I found this system easy to use.''
    (ii) ``I found it easy to follow the analysis flow.''
    
    \item \textbf{Transparency:}
    (i) ``I understood why the system reached its conclusion.''
    (ii) ``Intermediate steps (data selection, graph, estimation) were sufficiently explained.''
    (iii) ``It did not feel like the answer was produced suddenly without explanation.''
    
    \item \textbf{Trust:}
    (i) ``I would use the result of this task for real decision making.''
    (ii) ``When I doubted the result, the system provided clues to verify it.''
    
    \item \textbf{Workload / Mental Effort (reverse-coded):}
    ``I experienced high mental effort while completing this task.''
    
    \item \textbf{Sense of Agency:}
    (i) ``I felt that I made the core decisions in the analysis.''
    (ii) ``It felt more like we analyzed together rather than the system analyzing instead of me.''
\end{itemize}

\paragraph{Open-ended questions (optional)}
Participants answered:
(i) ``What was the most confusing moment during this task?''
(ii) ``Which feature was most helpful during this task?''
(iii) Additional comments.

\paragraph{Post-experiment comparative items}
Participants responded to comparative statements such as:
(i) ``ORCA helped me reach a conclusion faster than the Baseline.''
(ii) ``ORCA reduced trial-and-error (re-asking / re-trying) compared to the Baseline.''
(iii) ``ORCA's results felt more trustworthy than the Baseline.''
(iv) ``I could quickly understand what to do next even as a first-time user.''
(v) ``The learning cost to become proficient with ORCA was not high.''
Participants also selected common blocking points in ORCA (multi-select).

\subsubsection{Results}
\label{appendix:user_study:survey:results}

\begin{table}[t]
\centering
\begin{tabular}{lccc}
\hline
\textbf{Construct} & \textbf{Baseline} & \textbf{ORCA} & \textbf{$\Delta$} \\
\hline
Usability 
& 4.21 & 4.83 & +0.63 \\
Transparency 
& 3.28 & 5.22 & +1.94 \\
Trust 
& 3.33 & 5.13 & +1.79 \\
Mental Effort (R) 
& 3.75 & 4.00 & +0.25 \\
Sense of Agency 
& 3.63 & 4.58 & +0.96 \\
\hline
\end{tabular}
\caption{Post-task ratings aggregated by construct (7-point Likert).}
\label{tab:appendix_post_survey}
\end{table}

\paragraph{Post-task ratings}
Table~\ref{tab:appendix_post_survey} summarizes participants’ post-task ratings on a 7-point Likert scale.
Across nearly all dimensions, ORCA received higher average ratings than the Baseline.
The largest gains were observed in \textbf{Transparency}, where participants reported clearer explanations of intermediate steps and less perception of results appearing as a ``sudden output''.
Improvements were also observed in \textbf{Trust}, with participants indicating greater willingness to rely on results and better availability of clues for verification.
Usability and Sense of Agency showed moderate but consistent improvements, while reported mental effort (reverse-coded) was slightly lower for ORCA, suggesting that the added structure did not increase cognitive burden.

\begin{table}[t]
\centering
\begin{tabular}{lcc}
\hline
\textbf{Item} & \textbf{Mean} & \textbf{Std.} \\
\hline
ORCA helped me reach conclusions faster & 4.33 & 1.97 \\
ORCA reduced trial-and-error & 4.67 & 2.31 \\
ORCA results felt more trustworthy & 5.50 & 1.98 \\
ORCA was easy to orient as a first-time user & 4.67 & 1.56 \\
Learning cost to become proficient was low & 4.25 & 1.54 \\
\hline
\end{tabular}
\caption{Post-experiment survey (7-point Likert).}
\label{tab:appendix_post_comparison}
\end{table}

\paragraph{Post-experiment comparative ratings}
Table~\ref{tab:appendix_post_comparison} summarizes participants’ comparative judgments after completing both conditions.
Participants generally perceived ORCA as more trustworthy than the Baseline and reported improved clarity in understanding what to do next.
Perceived gains in efficiency, such as faster convergence and reduced trial-and-error, were positive on average but showed variability across participants.
Overall, these results suggest that ORCA’s primary perceived advantage lies in enhanced clarity and trustworthiness, with efficiency gains depending on individual user experience.

\newpage
\section{Agent Prompt Templates}
\label{appendix-sec:prompts}
\setcounter{figure}{0}
\setcounter{table}{0}
Below we list prompts for each module in ORCA.

\begin{figure*}[t]
    \centering
    \includegraphics[
        width=0.8\linewidth,
        keepaspectratio
    ]{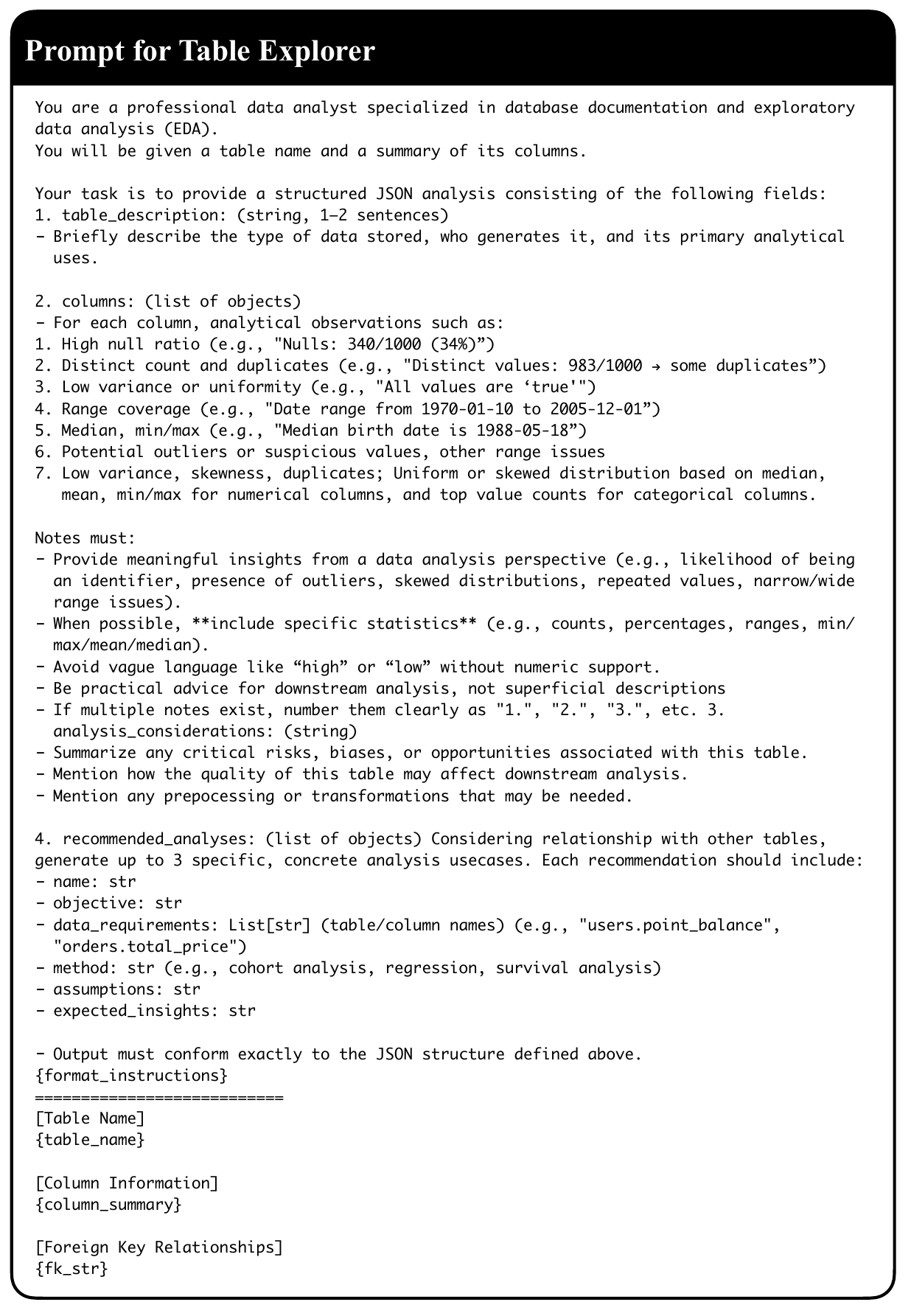}
    \caption{Prompt for Table Explorer module. Generates a JSON output including summary statistics, anomalies, and recommended analyses.}
    \Description{Prompt for Table Explorer module. Generates a JSON output including summary statistics, anomalies, and recommended analyses.}
    \label{fig:table_explorer_prompt}
\end{figure*}

\begin{figure*}[t]
    \centering
    \includegraphics[
        width=0.8\linewidth,
        keepaspectratio
    ]{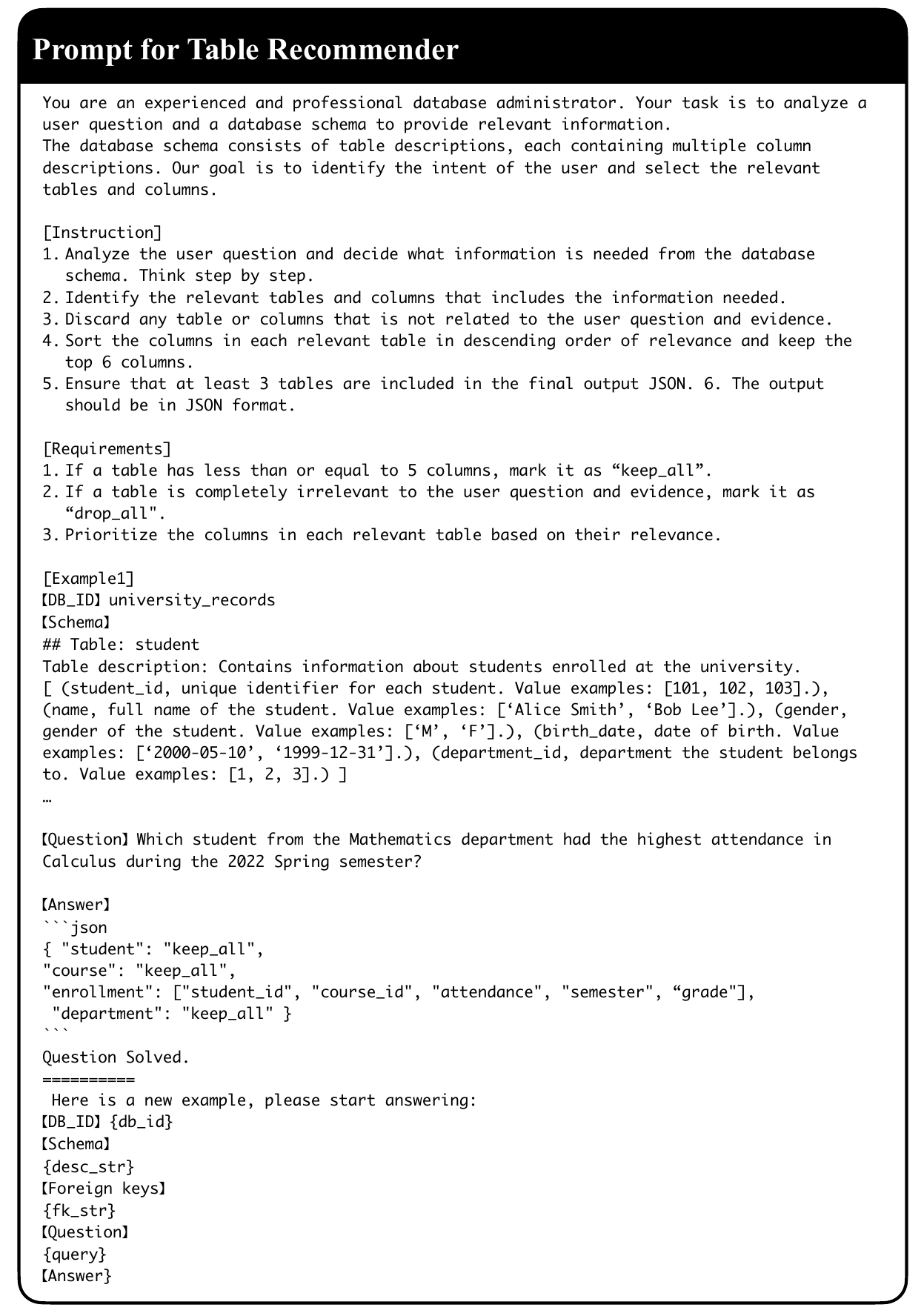}
    \caption{Prompt for Table Recommender module. Due to space constraints, only one of the few-shot prompt examples is included, with some parts omitted.}
    \Description{Prompt for Table Recommender module. Due to space constraints, only one of the few-shot prompt examples is included, with some parts omitted.}
    \label{fig:table_recommender_prompt}
\end{figure*}

\begin{figure*}[t]
    \centering
    \includegraphics[
        width=0.8\linewidth,
        keepaspectratio
    ]{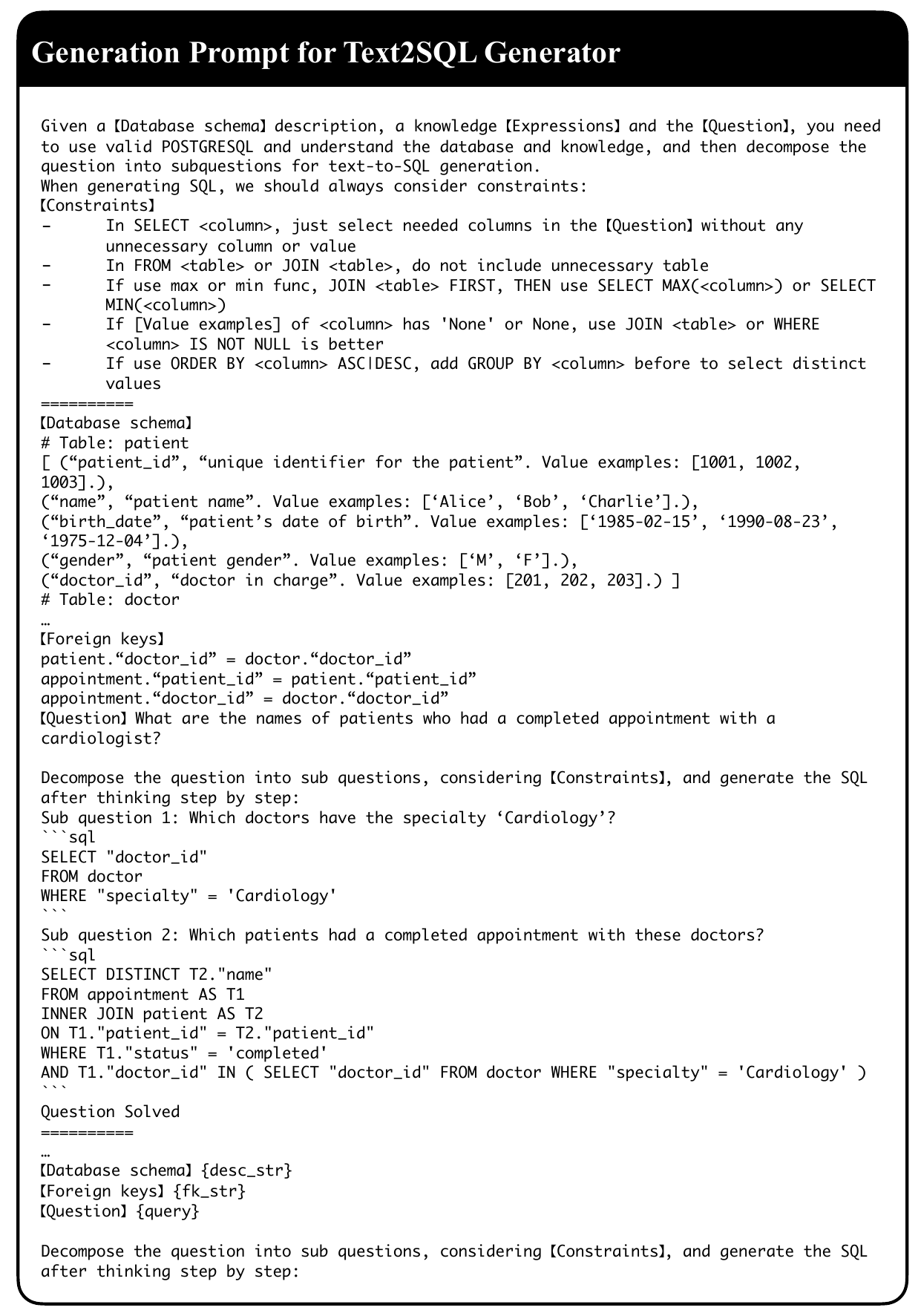}
    \caption{Generation prompt for Text2SQL Generator module. Due to space constraints, only one of the few-shot prompt examples is included, with some parts omitted.}
    \Description{Generation prompt for Text2SQL Generator module. Due to space constraints, only one of the few-shot prompt examples is included, with some parts omitted.}
    \label{fig:text2sql_generation_prompt}
\end{figure*}

\begin{figure*}[t]
    \centering
    \includegraphics[
        width=0.8\linewidth,
        keepaspectratio
    ]{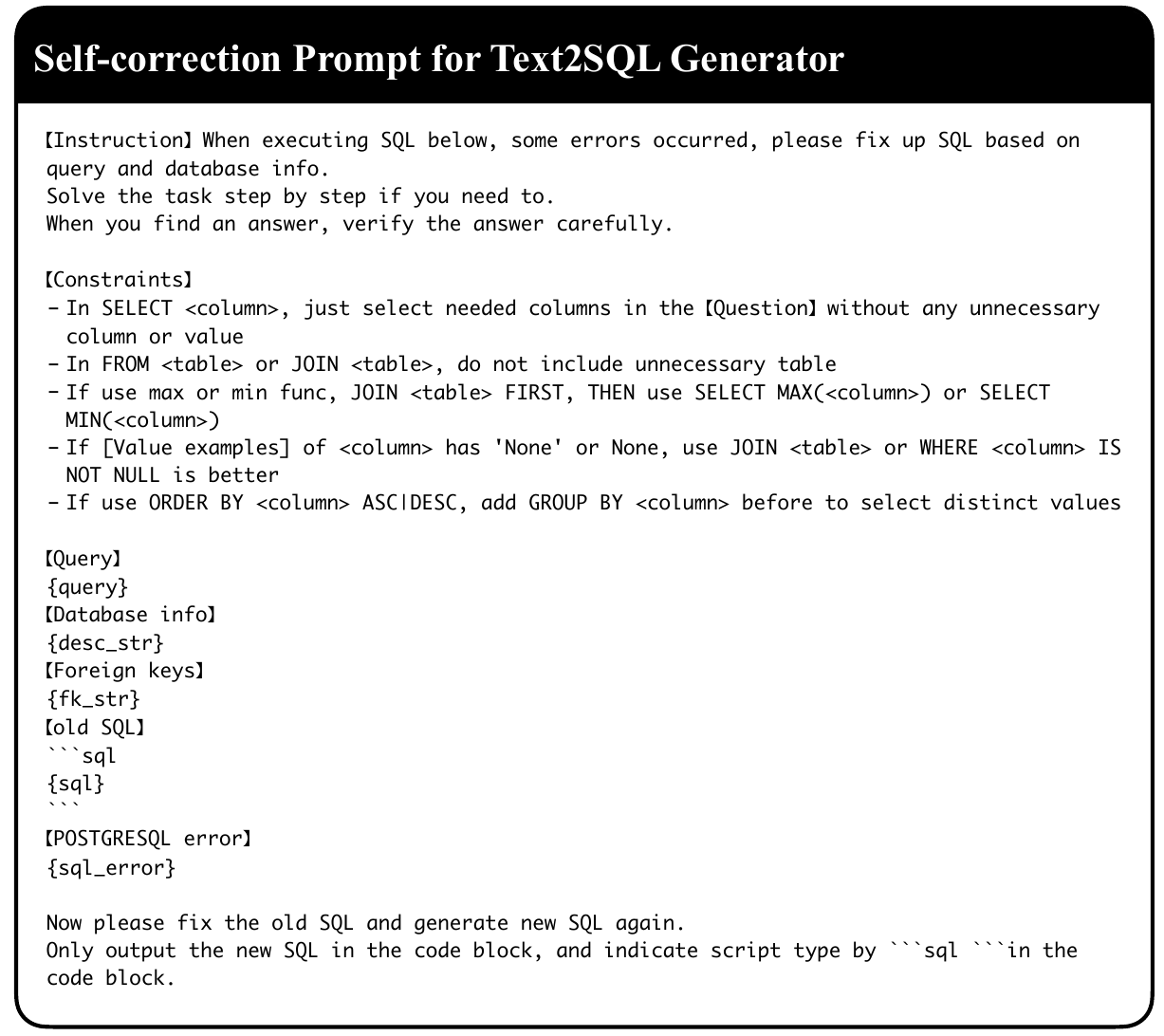}
    \caption{Self-correction prompt for Text2SQL Generator module, revises faulty SQL codes.}
    \Description{Self-correction prompt for Text2SQL Generator module, revises faulty SQL codes.}
    \label{fig:text2sql_selfcorrection_prompt}
\end{figure*}

\begin{figure*}[t]
    \centering
    \includegraphics[
        width=0.8\linewidth,
        keepaspectratio
    ]{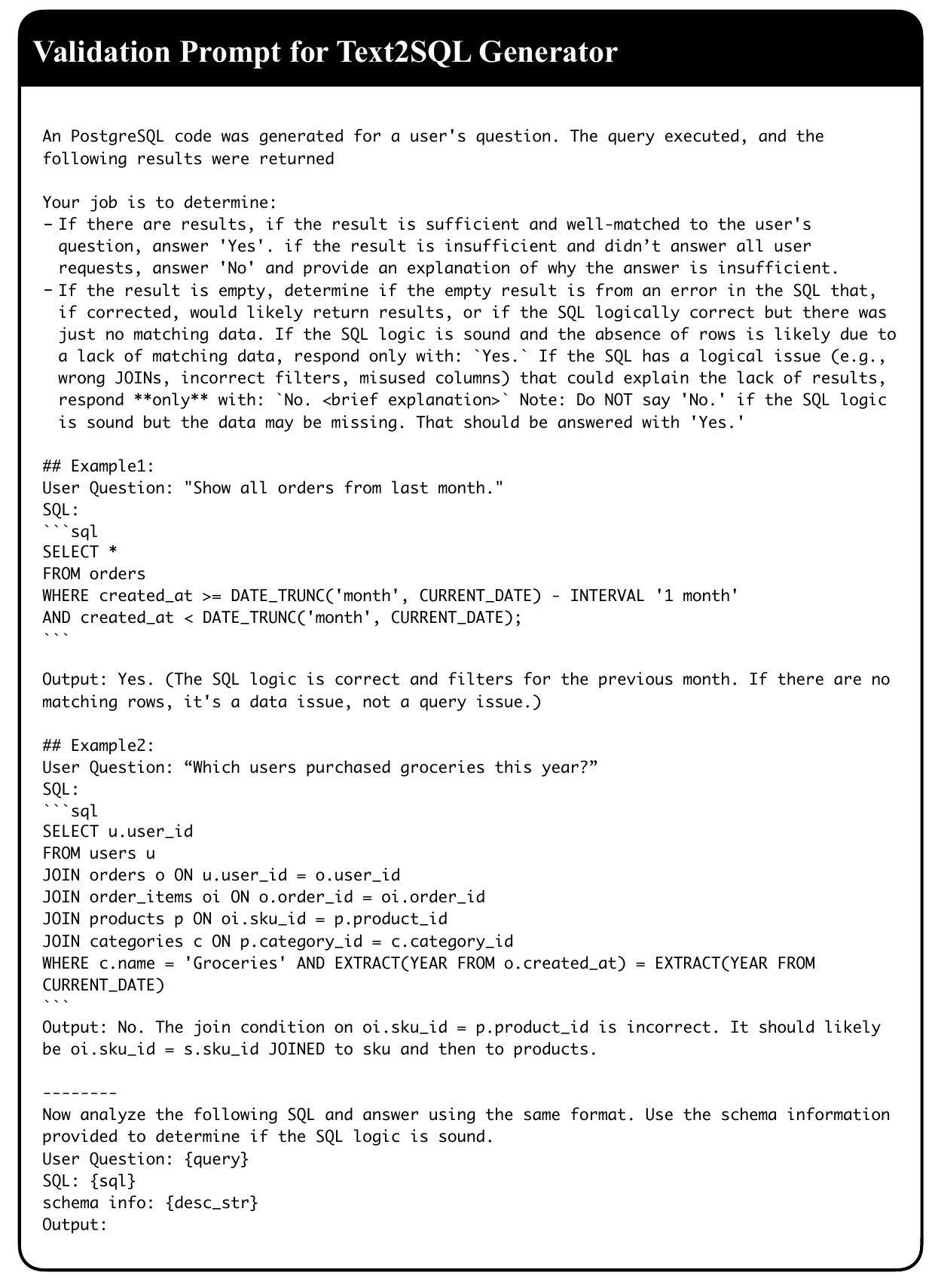}
    \caption{Validation prompt for Text2SQL Generator module, assessing whether the generated SQL code fulfills the user request and is free of logical errors.}
    \Description{Validation prompt for Text2SQL Generator module, assessing whether the generated SQL code fulfills the user request and is free of logical errors.}
    \label{fig:text2sql_validation_prompt}
\end{figure*}

\begin{figure*}[t]
    \centering
    \includegraphics[
        width=0.8\linewidth,
        keepaspectratio
    ]{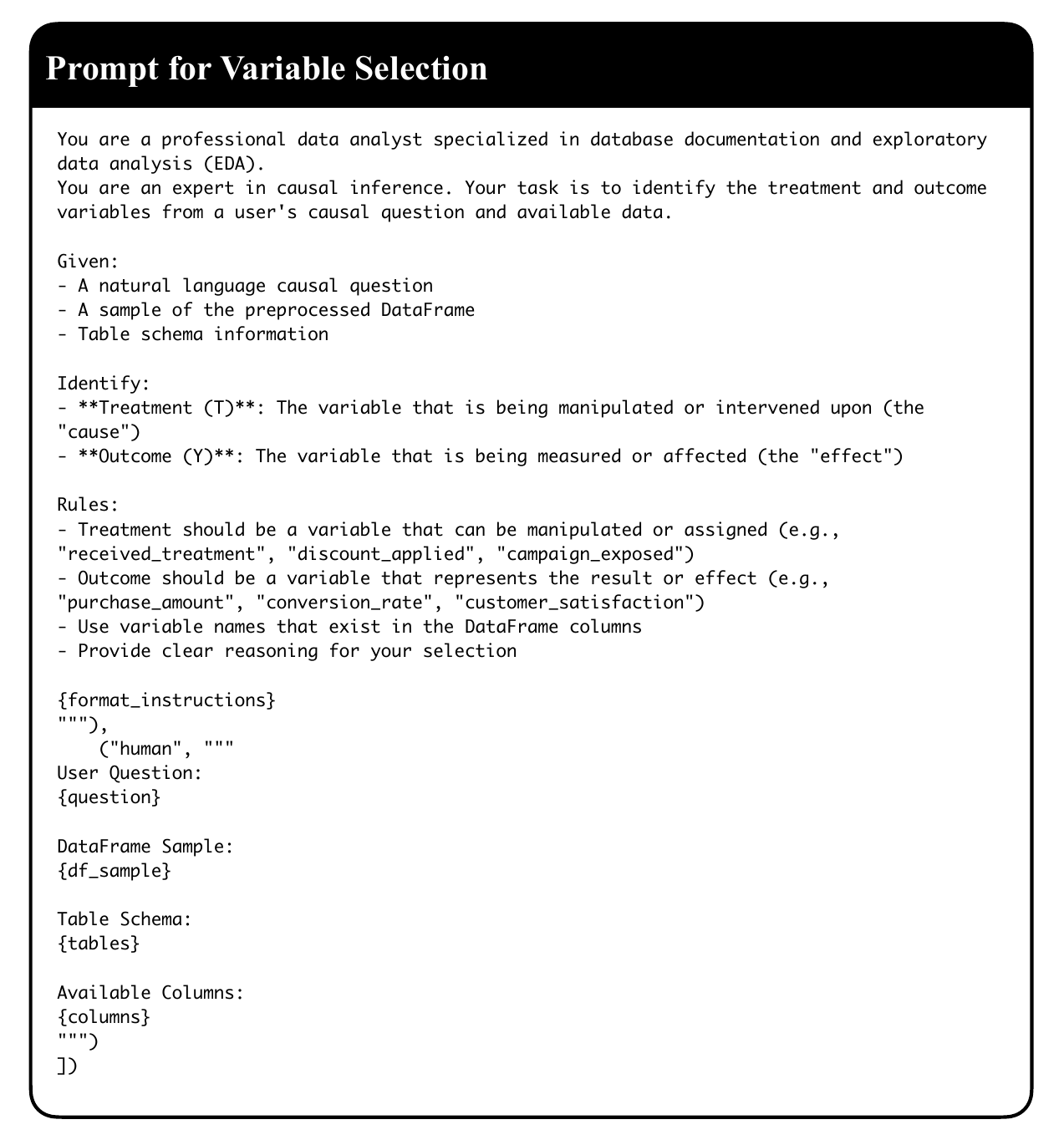}
    \caption{Prompt for the Variable Selection module of Causal Inference Agent, identifies the treatment and outcome variable based on user query, dataframe, and schema information.}
    \Description{Prompt for the Variable Selection module of Causal Inference Agent, identifies the treatment and outcome variable based on user query, dataframe, and schema information.}
    \label{fig:variable_selection_prompt}
\end{figure*}

\begin{figure*}[t]
    \centering
    \includegraphics[
        width=0.8\linewidth,
        keepaspectratio
    ]{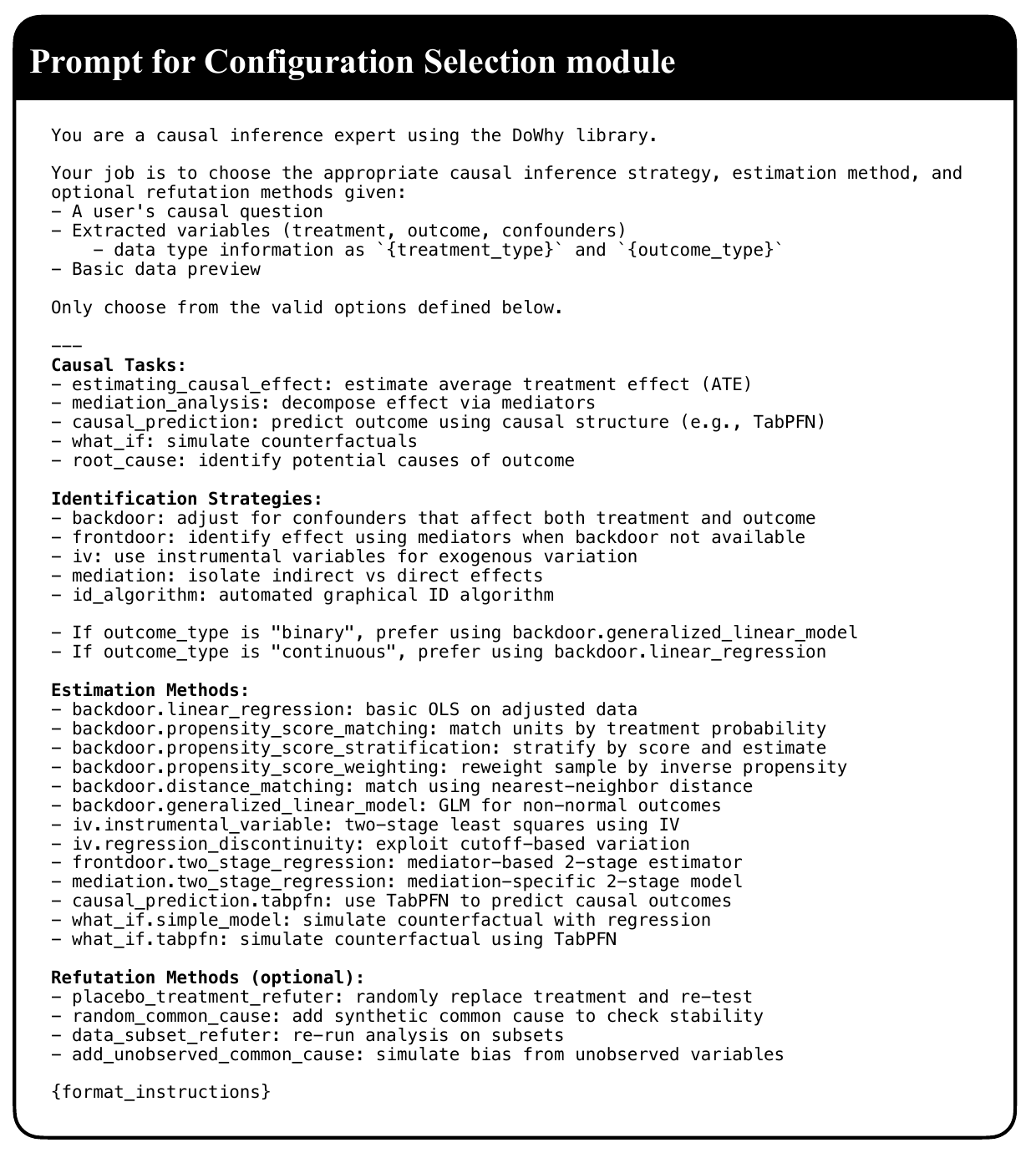}
    \caption{Prompt for Configuration Selection module, selects a causal inference strategy including the task, identification, estimation, and refutation based on user query and data context.}
    \Description{Prompt for Config Selector module, selects a causal inference strategy including the task, identification, estimation, and refutation based on user query and data context.}
    \label{fig:config_selector_prompt}
\end{figure*}

\begin{figure*}[t]
    \centering
    \includegraphics[
        width=0.8\linewidth,
        keepaspectratio
    ]{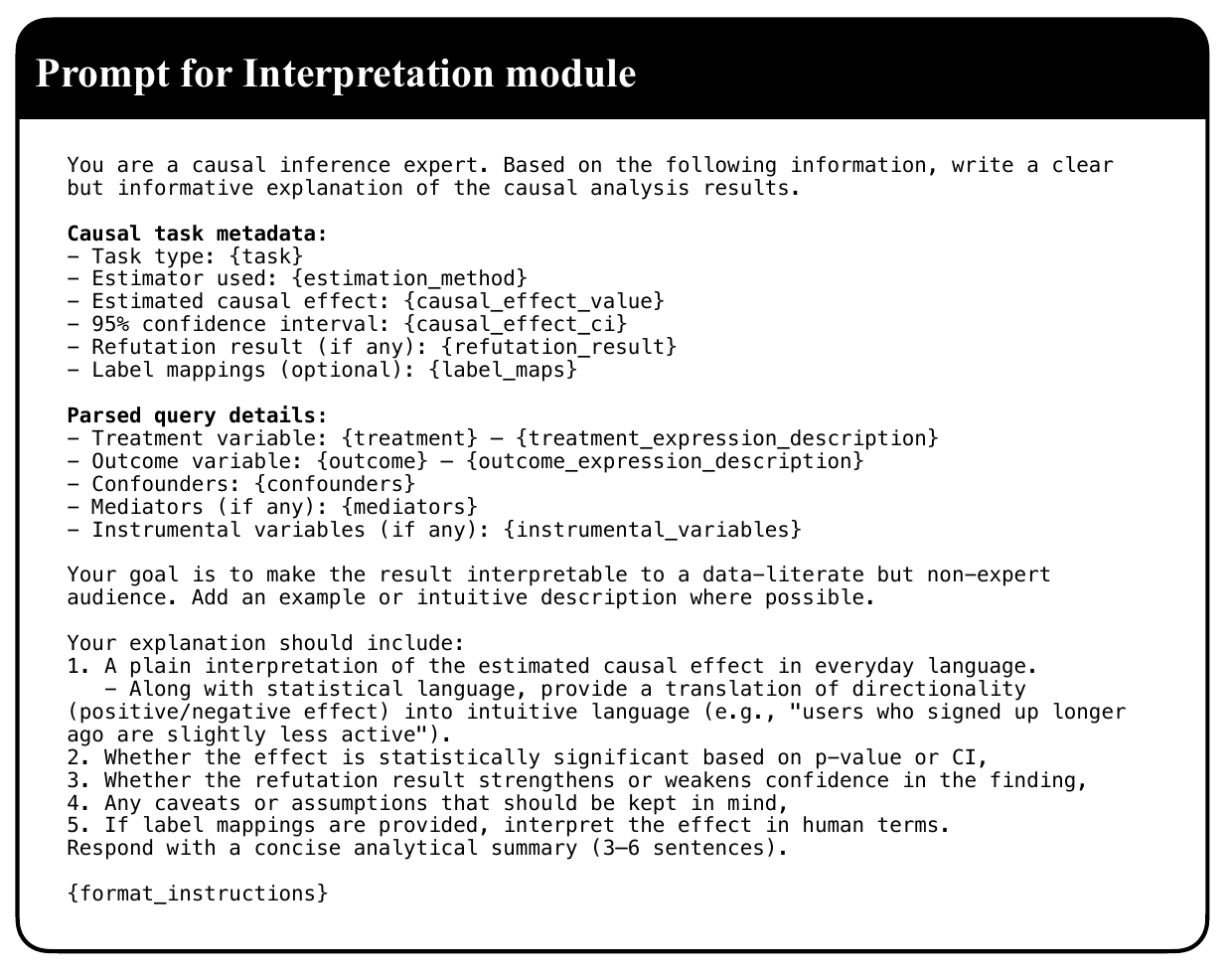}
    \caption{Prompt for the Interpretation module of Causal Inference Agent, guiding natural-language explanation of causal effect estimates and significance.}
    \Description{Prompt for the Interpretation module of Causal Inference Agent, guiding natural-language explanation of causal effect estimates and significance.}
    \label{fig:interpretation_prompt}
\end{figure*}

\begin{figure*}[t]
    \centering
    \includegraphics[
        width=0.8\linewidth,
        keepaspectratio
    ]{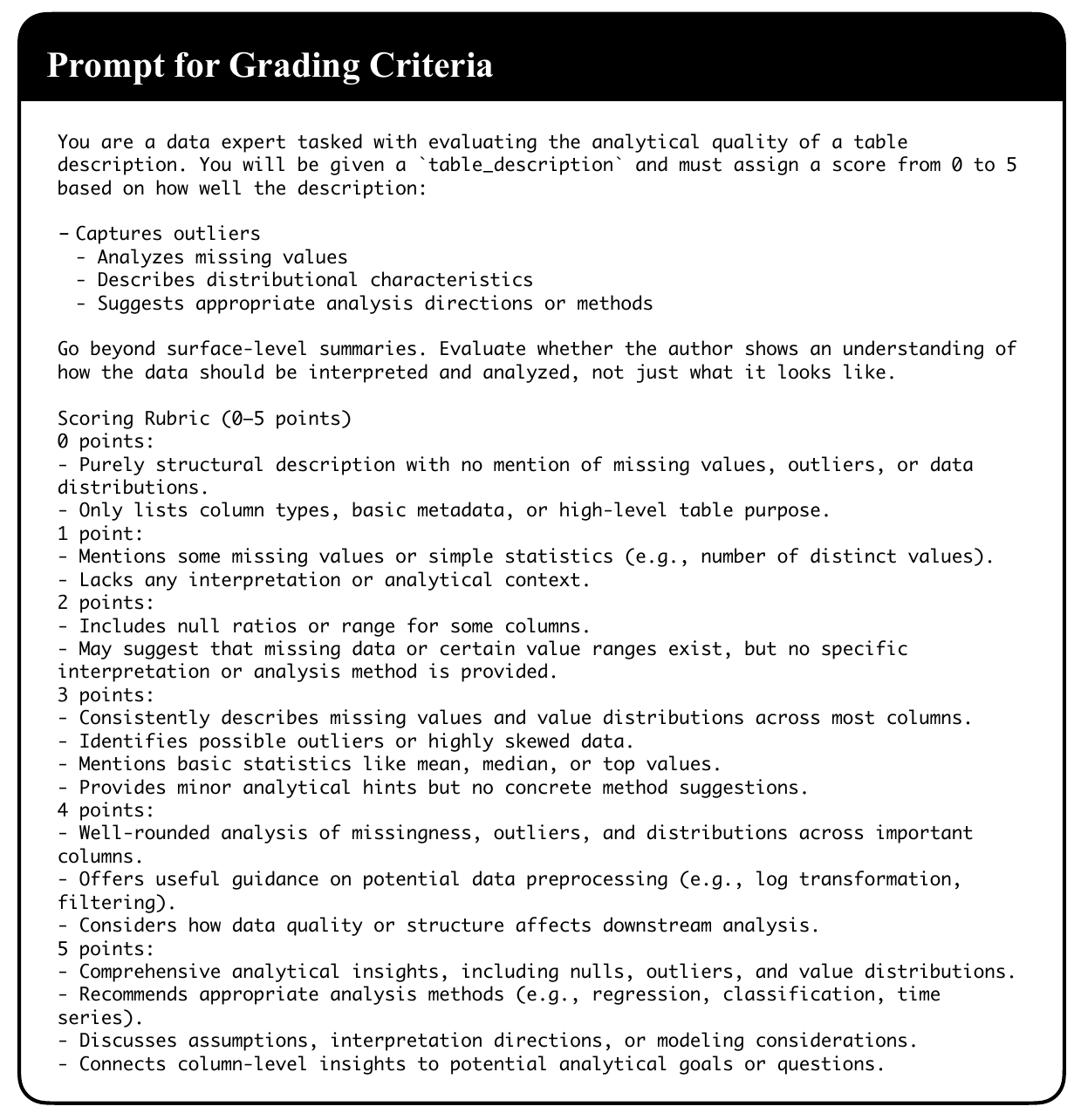}
    \caption{Evaluation Prompt for Table Description Task.  
    GPT-4o mini use this prompt for scoring table descriprion of table explorer module and baseline(GPT-4o-mini).}
    \Description{Evaluation Prompt for Table Description Task.  
    GPT-4o mini use this prompt for scoring table descriprion of table explorer module and baseline(GPT-4o-mini).}
    \label{fig:table_description_criteria}
\end{figure*}

\clearpage

\begin{figure*}[t]
    \centering
    \includegraphics[
        width=0.8\linewidth,
        keepaspectratio
    ]{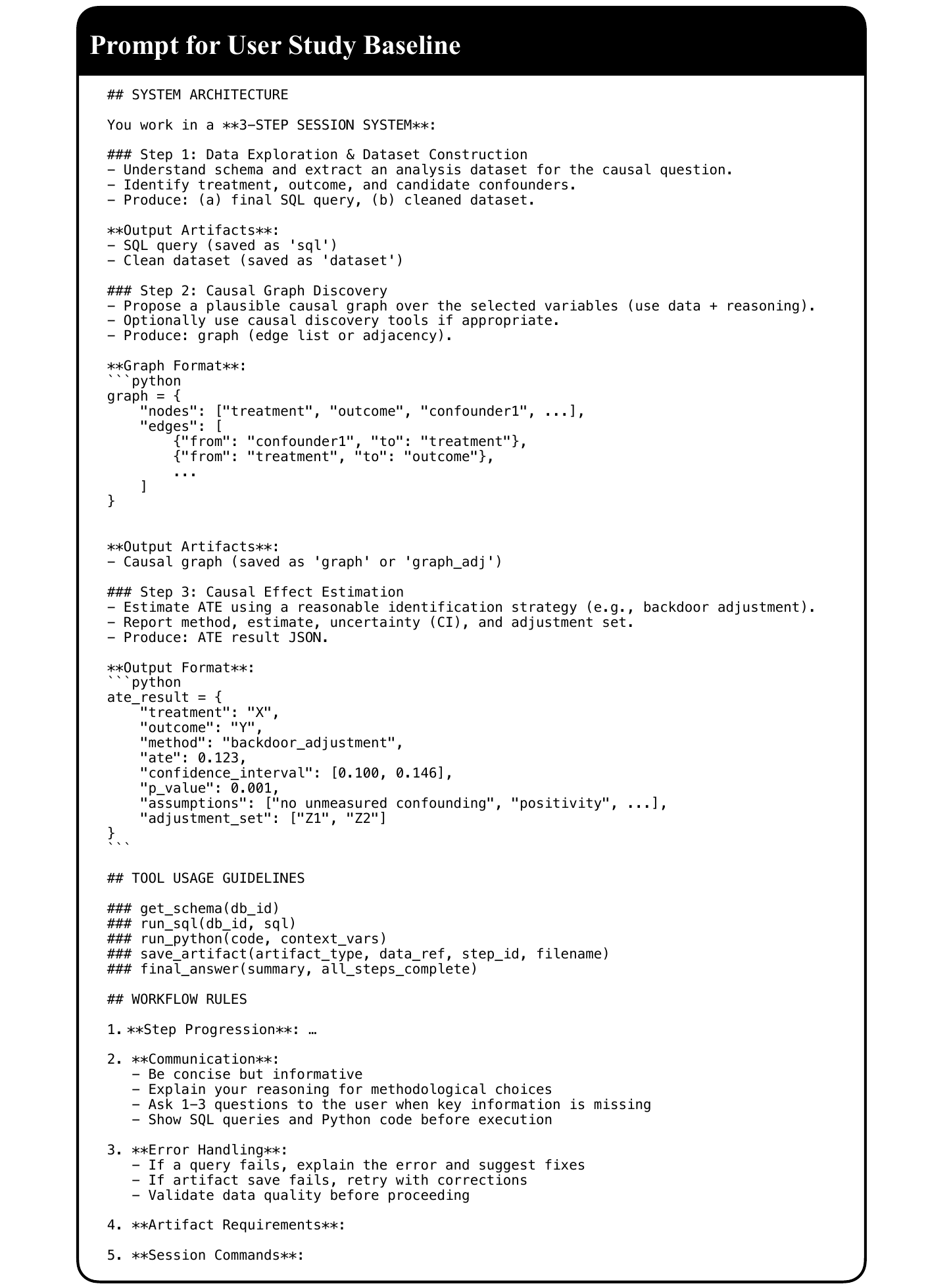}
    \caption{Prompt for the user study baseline LLM-assistant with a step-gated workflow for end-to-end causal analysis.
    The prompt enforces a three-step session structure (data retrieval, causal graph specification, and effect estimation), explicit step transitions, and required artifact saving at each stage, while allowing free-form interaction and tool use within steps.}
    \Description{Prompt for the user study baseline LLM-assistant with a step-gated workflow for end-to-end causal analysis.
    The prompt enforces a three-step session structure (data retrieval, causal graph specification, and effect estimation), explicit step transitions, and required artifact saving at each stage, while allowing free-form interaction and tool use within steps.}
    \label{fig:baseline_prompt}
\end{figure*}

\clearpage

\section{Sample Output of modules}
\label{appendix-sec:sample-log}
\setcounter{figure}{0}
\setcounter{table}{0}
This section provides example outputs generated by ORCA modules.

\begin{figure*}[t]
    \centering
    \includegraphics[width=0.95\linewidth]{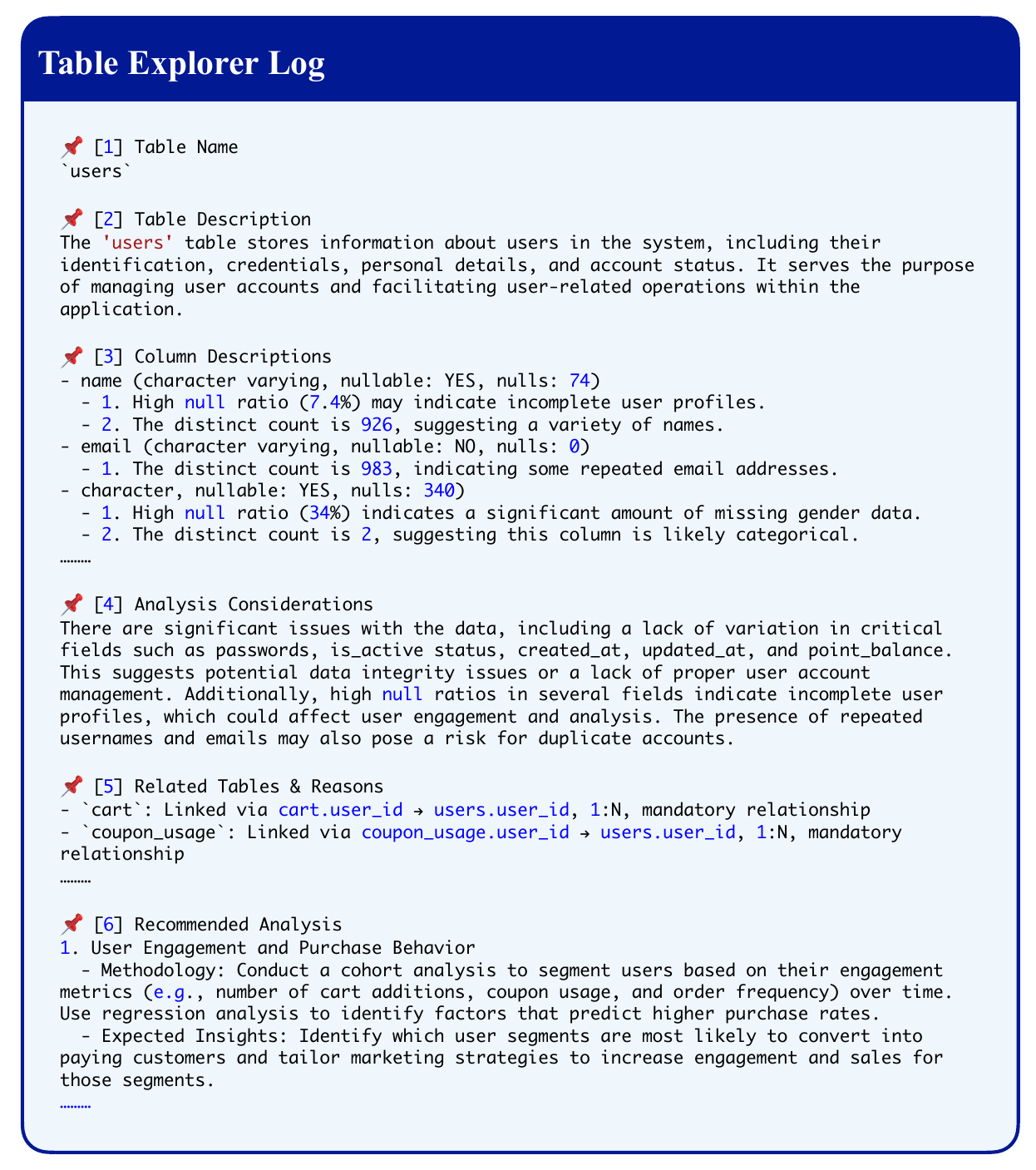}
    \caption{Execution log of the Table Explorer module, describing a given table by highlighting column specific characteristics, providing relational context and suggesting analysis considerations.}
    \Description{Execution log of the Table Explorer module, describing a given table by highlighting column specific characteristics, providing relational context and suggesting analysis considerations.}
    \label{fig:table-explorer-log}
\end{figure*}

\begin{figure*}[t]
    \centering
    \includegraphics[width=0.95\linewidth]{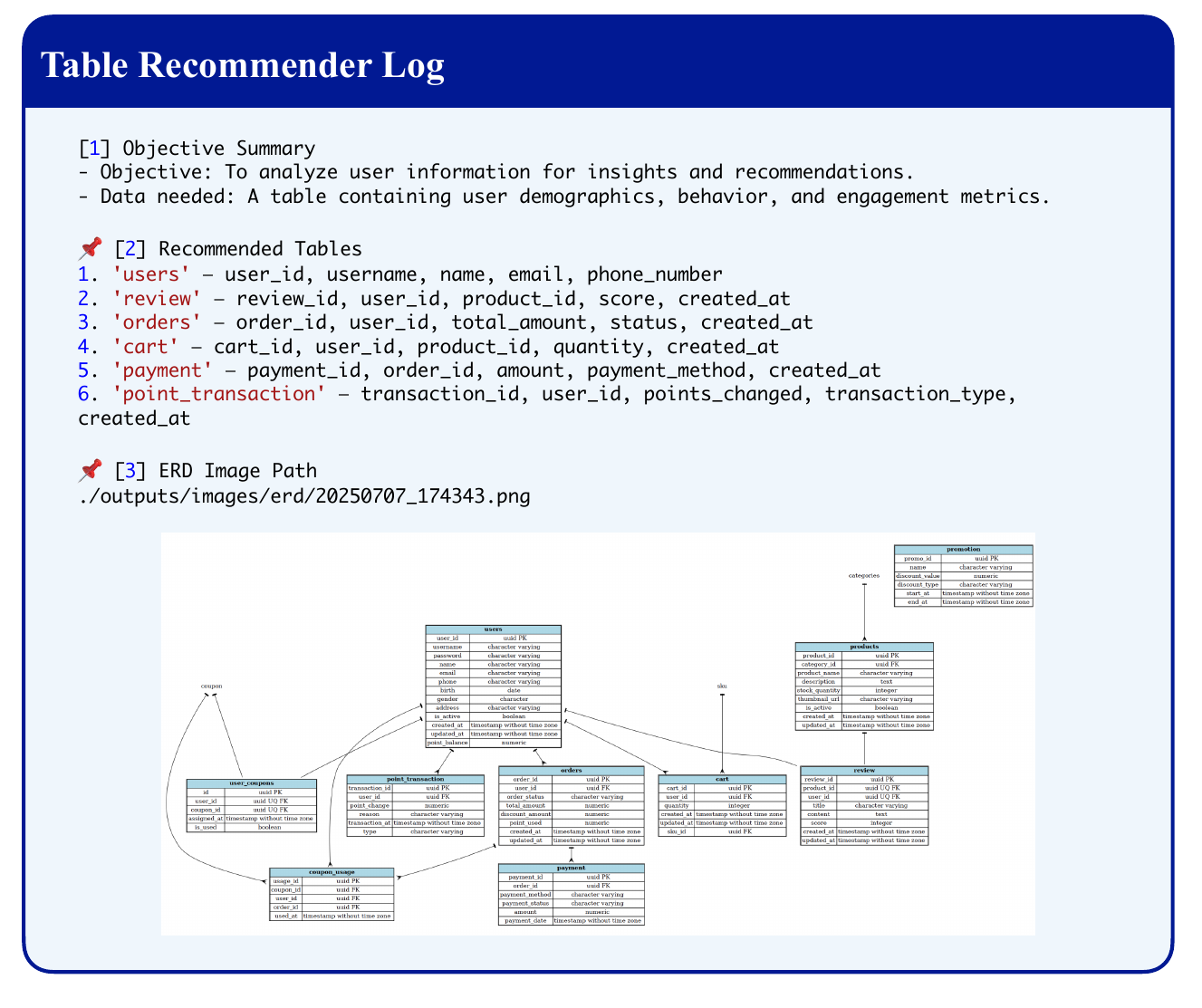}
    \caption{Execution log of the Table Recommender module, presenting the tables and columns most relevant to the analysis, along with a visualization of their relationships.}
    \Description{Execution log of the Table Recommender module, presenting the tables and columns most relevant to the analysis, along with a visualization of their relationships.}
    \label{fig:table-rec-log}
\end{figure*}

\begin{figure*}
    \centering
    \includegraphics[width=0.95\linewidth]{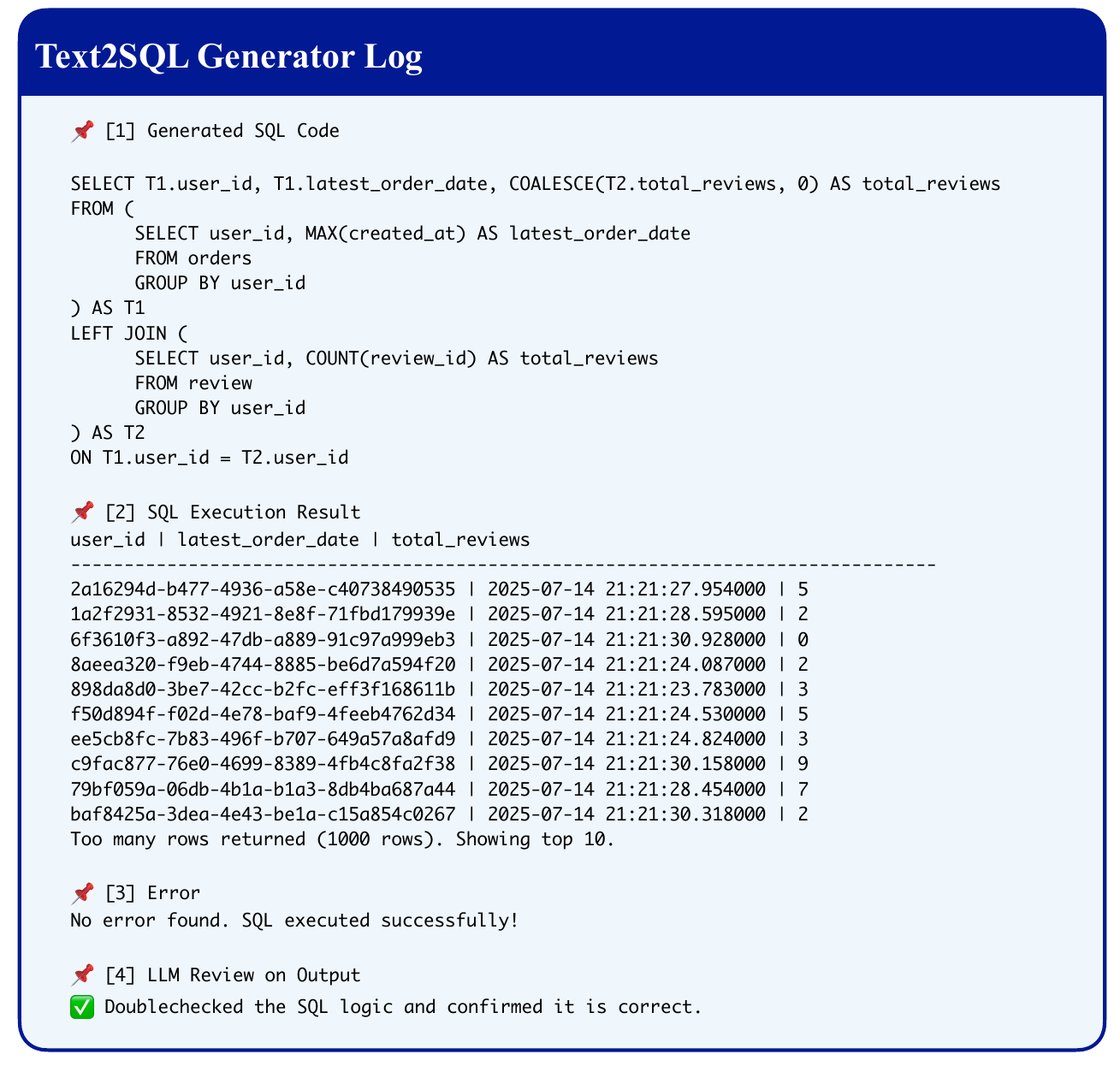}
    \caption{Execution log of the Text2SQL Generator module, displaying the generated SQL, partial results, encountered errors (if any), and module feedback}
    \Description{Execution log of the Text2SQL Generator module, displaying the generated SQL, partial results, encountered errors (if any), and module feedback}
    \label{fig:text2sql-generator-log}
\end{figure*}

\begin{figure*}
    \centering
    \includegraphics[
        width=0.8\linewidth,
        keepaspectratio
    ]{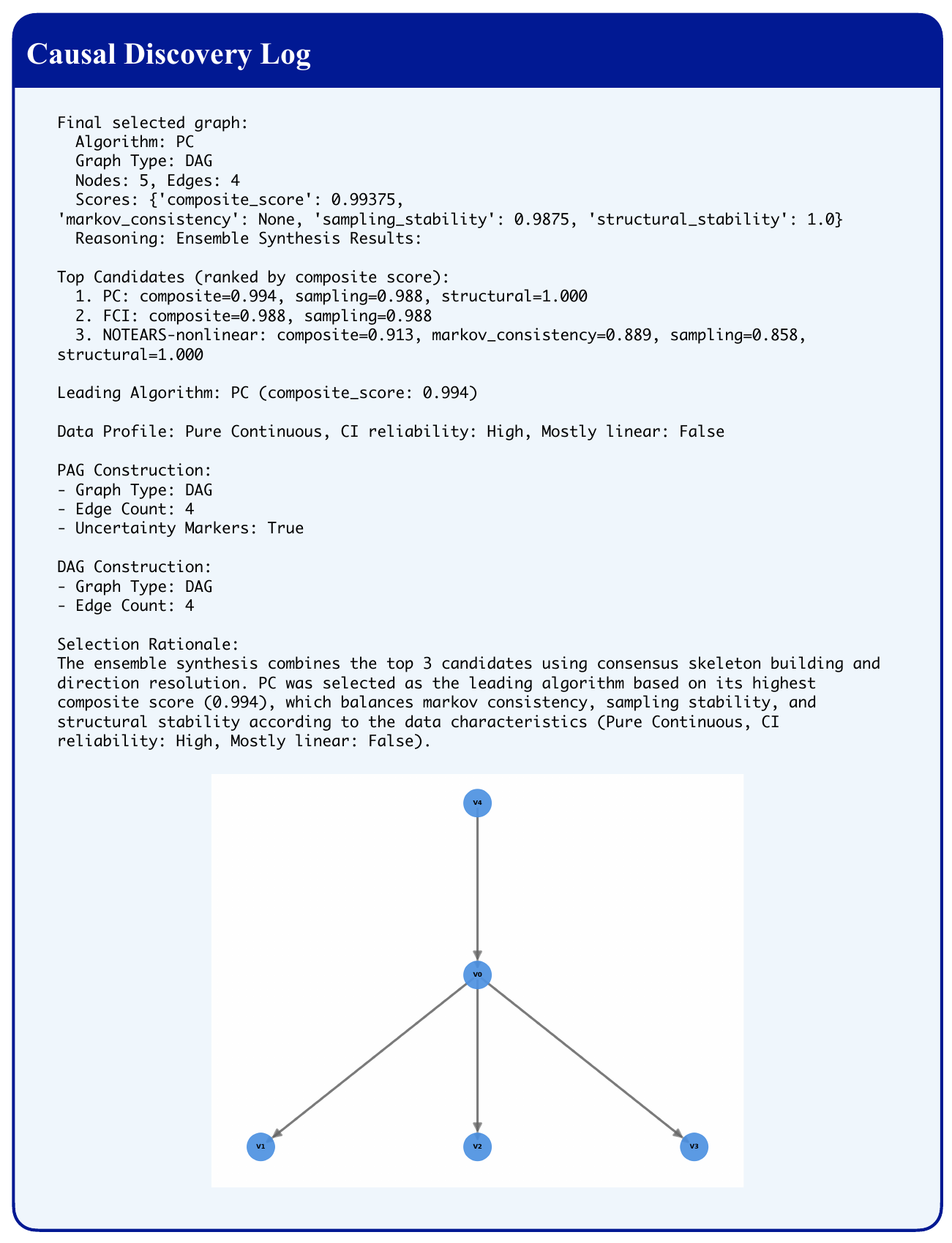}
    \caption{Execution log of the Causal Discovery Agent. It summarizes candidate graphs produced by multiple discovery algorithms, their composite scores, inferred data characteristics, and the rationale for selecting the final graph. In this example, all algorithms return fully oriented DAGs, so the selected DAG is visualized.}
    \Description{Execution log of the Causal Discovery Agent. It summarizes candidate graphs produced by multiple discovery algorithms, their composite scores, inferred data characteristics, and the rationale for selecting the final graph. In this example, all algorithms return fully oriented DAGs, so the selected DAG is visualized.}
    \label{fig:discovery-log}
\end{figure*}

\begin{figure*}
    \centering
    \includegraphics[
        width=0.8\linewidth,
        keepaspectratio
    ]{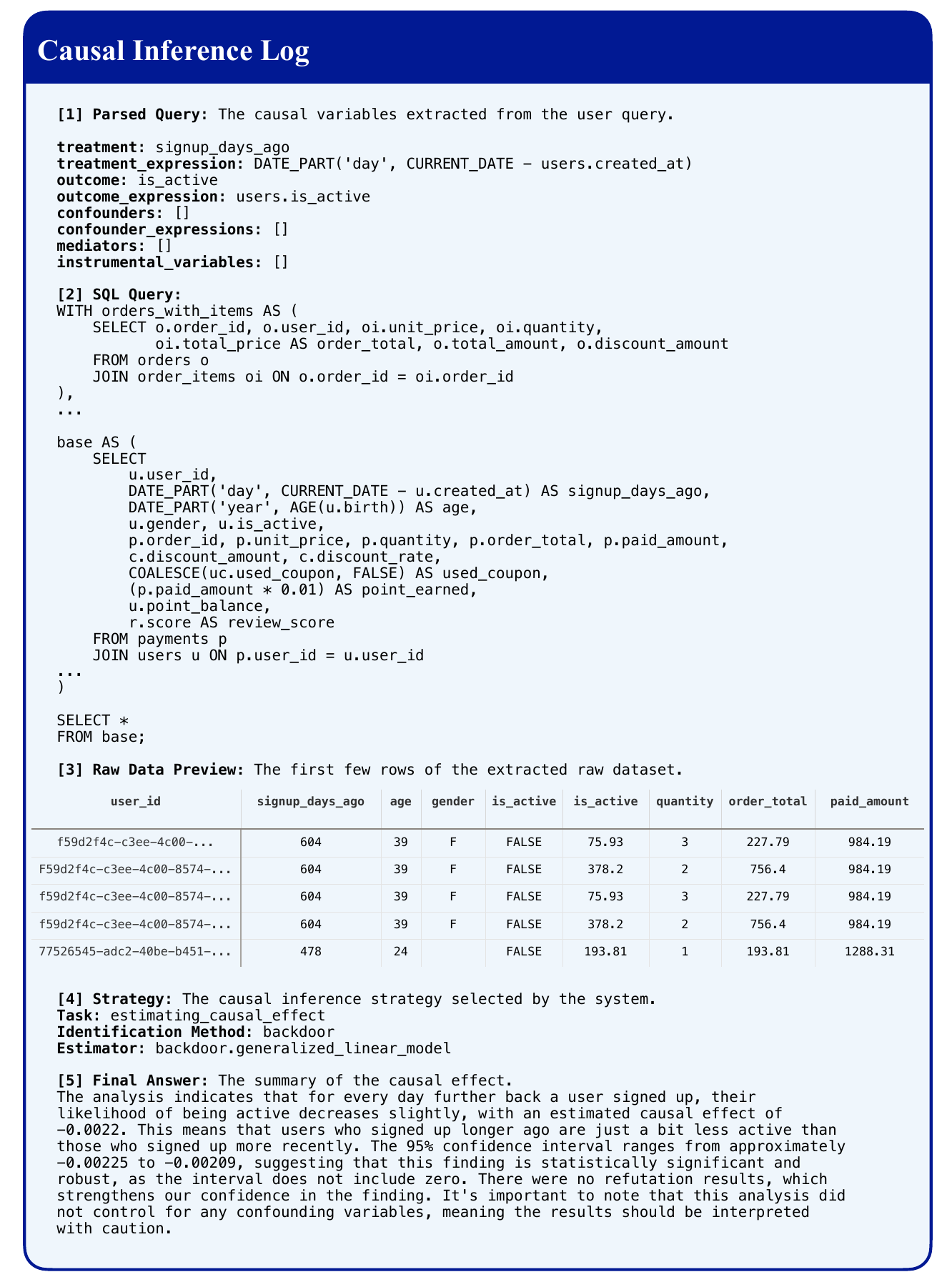}
    \caption{Execution log of the Causal Inference Agent. It includes the information used through the analysis such as parsed user query, preview of the data, a causal inference strategy, along with an interpretable summary of the estimation result.}
    \Description{Execution log of the Interpretation module. It includes the information used through the analysis such as parsed user query, preview of the data, a causal inference strategy, along with an interpretable summary of the estimation result.}
    \label{fig:analyzer-log}
\end{figure*}

\clearpage

\end{document}